\documentclass[prx,twocolumn,superscriptaddress]{revtex4-1}
\usepackage[english]{babel}
\usepackage[dvips]{graphicx}
\usepackage{latexsym}
\usepackage{amsthm}
\usepackage{amsfonts}
\usepackage{amssymb}

\usepackage[colorlinks=true,citecolor=blue,linkcolor=red,urlcolor=blue, linktocpage=true,breaklinks=true,pagebackref=false]{hyperref}

\usepackage{textcomp} 		
\usepackage{hyperref} 		
\usepackage{ifthen} 			
\usepackage{xifthen} 		
\usepackage{color} 			
\usepackage{soul} 			
\usepackage{amsmath} 		
\usepackage{lineno}
\usepackage[normalem]{ulem}

\usepackage{xcolor} 
\usepackage{float}

\newcommand{\uproman}[1]{\uppercase\expandafter{\romannumeral#1}}

\begin{document}


\title{Spectral evidence of squeezing of a weakly damped driven nanomechanical mode}
\author{J.\, S. Huber}
\affiliation{Department of Physics, University of Konstanz, 78457 Konstanz, Germany}
\author{G. Rastelli}
\affiliation{Department of Physics, University of Konstanz, 78457 Konstanz, Germany}
\author{M.\, J. Seitner}
\affiliation{Department of Physics, University of Konstanz, 78457 Konstanz, Germany}
\author{J.\, K\"olbl}
\altaffiliation[Present address: ]{Department of Physics, University of Basel, 4056 Basel, Switzerland}
\affiliation{Department of Physics, University of Konstanz, 78457 Konstanz, Germany}
\author{W. Belzig}
\affiliation{Department of Physics, University of Konstanz, 78457 Konstanz, Germany}
\author{M.\,I. Dykman}
\altaffiliation{dykmanm@msu.edu}
\affiliation{Michigan State University, East Lansing, MI 48824, USA}
\author{E.\,M. Weig}
\altaffiliation{eva.weig@uni-konstanz.de}
\affiliation{Department of Physics, University of Konstanz, 78457 Konstanz, Germany}

\begin{abstract} 
Because of the broken time-translation symmetry, in periodically driven vibrational systems fluctuations of different vibration components have different intensities. Fluctuations of one of the components are often squeezed, whereas fluctutions of the other component, which is shifted in phase by $\pi/2$, are increased. 
Squeezing is a multi-faceted phenomenon, it attracts much attention from the perspective of high-precision measurements.
Here we demonstrate a new and hitherto unappreciated side of squeezing: its direct manifestation in the spectra of driven vibrational systems.
With a weakly damped nanomechanical resonator, we study the spectrum of thermal fluctuations of a resonantly driven nonlinear mode. 
In the attained sideband-resolved regime, we show that the asymmetry of the spectrum directly characterizes the squeezing. 
This opens a way to deduce squeezing of thermal fluctuations in strongly underdamped resonators, for which a direct determination by a standard homodyne measurement is impeded by frequency fluctuations.
The experimental and theoretical results are in excellent agreement. 
We further extend the theory to also describe the spectral manifestation of squeezing of quantum fluctuations.

\end{abstract}

\date{\today}

\maketitle

%
%
%
%
\section{Introduction} 

When appropriately scaled, the coordinate and momentum of a vibrational system or their canonical conjugate linear combinations form two vibration components. The scaling is done in such a way that, classically, the components oscillate with equal amplitudes in an isolated system, whereas their phases differ by $\pi/2$. If the system is coupled to a thermal reservoir, the vibration components fluctuate with the same intensities, in the absence of driving. This is a consequence of the time translation symmetry, as incrementing the time by a quarter of the oscillation period leads to the interchange of the components (modulo the sign). A periodic driving lifts the symmetry and can result in a reduction of fluctuations of one of the components, the effect of squeezing. Historically, squeezing was first detected in quantum optics \cite{Slusher:1985iw}. It attracted significant attention, since it can reduce the fluctuations below the quantum limit imposed by the uncertainty principle in the absence of driving \cite{Walls:1983ei}. This enables high-precision measurements \cite{Caves1981,Aasi2013,LIGO2011,Tse2019,Acernese2019,Malnou:2019}. 
More recently, squeezing in the quantum regime was also demonstrated in mechanical systems \cite{Wollman2015,Lecocq2015,Pirkkalainen2015}.

However, the concept of squeezing of fluctuations in vibrational systems equally applies to the classical regime. Classical squeezing promises to reduce heating in computers \cite{Klaers2019a}; it also represents an important asset for high-precision sensing \cite{DiFilippo1992,Natarajan1995,Szorkovszky2013} and thus  paves the way for a new generation of nanomechanical detectors at room temperature.

Squeezing has been frequently accomplished using parametric pumping or radiation pressure and 
has been demonstrated and theoretically analyzed for microwave \cite{Yurke:1988ih,Yurke:1989ev} 
and mechanical \cite{Rugar:1991he,Carr:2000fo,Suh2010,Vinante2013,Szorkovszky2013,Poot2014,Sonar2018} 
resonators as well as for ions in a Penning trap \cite{Natarajan1995}.
The classical two-mode squeezing of mechanical resonators by non-degenerate parametric amplification has been also 
reported \cite{Mahboob:2014he,Patil2015,Pontin2016}.

Along with parametric oscillators, the other vibrational system intensely studied in different areas, from optics to circuit quantum electrodynamics and to nano- and micromechanical systems is the Duffing (Kerr) oscillator \cite{Dykman:2012}. This is an oscillator with quartic nonlinearity in the potential. When driven by a resonant field, it can display bistability of forced vibrations. From the time-symmetry argument,  one would expect the possibility of fluctuation squeezing in the corresponding vibrational states. A theory of the squeezing was developed in Ref.~\onlinecite{Buk-Yurke-PRE2006}. However, to date, squeezing in this system has been observed only in a narrow parameter range where the oscillator (a nanomechanical mode) was close to the cusp bifurcation point at which 
the branches of the stable vibrational states merge \cite{Almog:2007jz}. 

Even though squeezing is a feature of one of the vibration components, a natural question is whether the  decrease of fluctuations of a particular component is the only manifestation of squeezing. In our experiment we demonstrate that this is not the case. We reveal a different manifestation of squeezing and use it to characterize the squeezing quantitatively. The results demonstrate, in particular, that a driven Duffing oscillator displays strong squeezing in a broad parameter range.

Our approach is based on measuring the spectrum of a resonantly driven vibrational system. The spectra of fluctuations of such a system and of its response to an additional weak field display sideband peaks \cite{Dykman:1979,Drummond:1980c,Dykman:1994gl}. Such peaks  are separated from the peak at the strong-drive frequency and have been seen in micromechanical systems \cite{Stambaugh:2006kg}. If the vibrational system is strongly underdamped, the peaks are well-resolved. They come from the fluctuations of the amplitude and phase of forced vibrations about  their stable values determined by the drive and should have different heights and areas. The asymmetry of the spectrum  has been predicted to directly reflect the squeezing \cite{Dykman2011,Dykman2012,Andre2012}. 

In what follows we describe the observation of the sideband-resolved peaks in the fluctuation spectrum of a weakly damped driven nanomechanical resonator. Under sufficiently strong driving the spectrum shows  two perfectly resolved peaks symmetrically located on the opposite sides of the driving frequency, but indeed having different heights and areas.
The fluctuations are thermal, the resonant periodic drive is the only cause of the system being away from thermal equilibrium. In contrast to the previous experiments, no extra noise or extra drive is added. 
We use the asymmetry of the spectrum to infer the squeezing and determine the squeezing parameter. Our experimental results agree, with no adjustable parameters, with a theoretical model which extends the  
one discussed in \cite{Dykman:1979,Drummond:1980c,Dykman:1994gl,Dykman2012}.

It is instructive to compare our method with the conventional measurement of a squeezed state. The latter involves the measurement of the individual components (quadratures) of the vibrations, 
which is accomplished by controlling the phase between the vibrations and an injected signal. 
The commonly employed method to detect squeezing is a homodyne measurement. 
This technique has been used in all previous demonstrations of quantum or classical noise squeezing we are aware of, 
be it  the case of a parametric amplifier or  a Duffing resonator \cite{Slusher:1985iw,Wollman2015,Lecocq2015,Pirkkalainen2015,Yurke:1988ih,Yurke:1989ev,Rugar:1991he,Carr:2000fo,Suh2010,Vinante2013,Szorkovszky2013,Poot2014,Sonar2018,Natarajan1995,Mahboob:2014he,Patil2015,Pontin2016,Almog:2007jz}. 
In contrast, our method does not require measuring the individual vibrational components and does not involve homodyne detection. Rather it relies on the simple standard technique of spectral measurements. 
This is particularly favourable for strongly underdamped resonators such as the one explored in the present work, as the power spectrum is insensitive to frequency fluctuations as long as they are smaller than the decay rate, while the noise quadratures of weakly damped resonators are difficult to measure independently because of the accumulative effect of frequency fluctuations~\cite{Fong2012}. To the best of our knowledge, no homodyne measurement of single-mode squeezing of a strongly underdamped mechanical resonator has been reported. The advantage of measuring the power spectrum is not limited to mechanical resonators.

We also use our driven resonator to explore another effect that occurs in nonequilibrium systems with coexisting stable states.  For an equilibrium dynamical system such states can be thought of as the minima of a potential in which the system moves. Fluctuations cause switching between the states, forming a distribution over them. For a small fluctuation intensity, the state populations are exponentially different: in an equilibrium system, this difference is given by the Boltzmann factor that contains the difference between the potential minima divided by $k_BT$. Only in a narrow range where the minima are of almost equal depth are they almost equally populated, an analog of the first order phase transition. 

Generically, a nonequilibrium system does not have detailed balance and cannot be mapped onto a Brownian particle in a potential well. Still, it can display an analog of a kinetic phase transition where the state populations are almost equal \cite{Bonifacio1978, Dykman:1979}. A resonantly driven bistable classical oscillator is a system for which it was predicted where such a transition occurs \cite{Dykman:1979}. Our nanoresonator allows us to find the kinetic phase transition in a system lacking detailed balance and thus to quantitatively test a major aspect of the theory of fluctuations in such systems.

%
%
%
%
\begin{figure*}[t]
\includegraphics[scale=0.12]{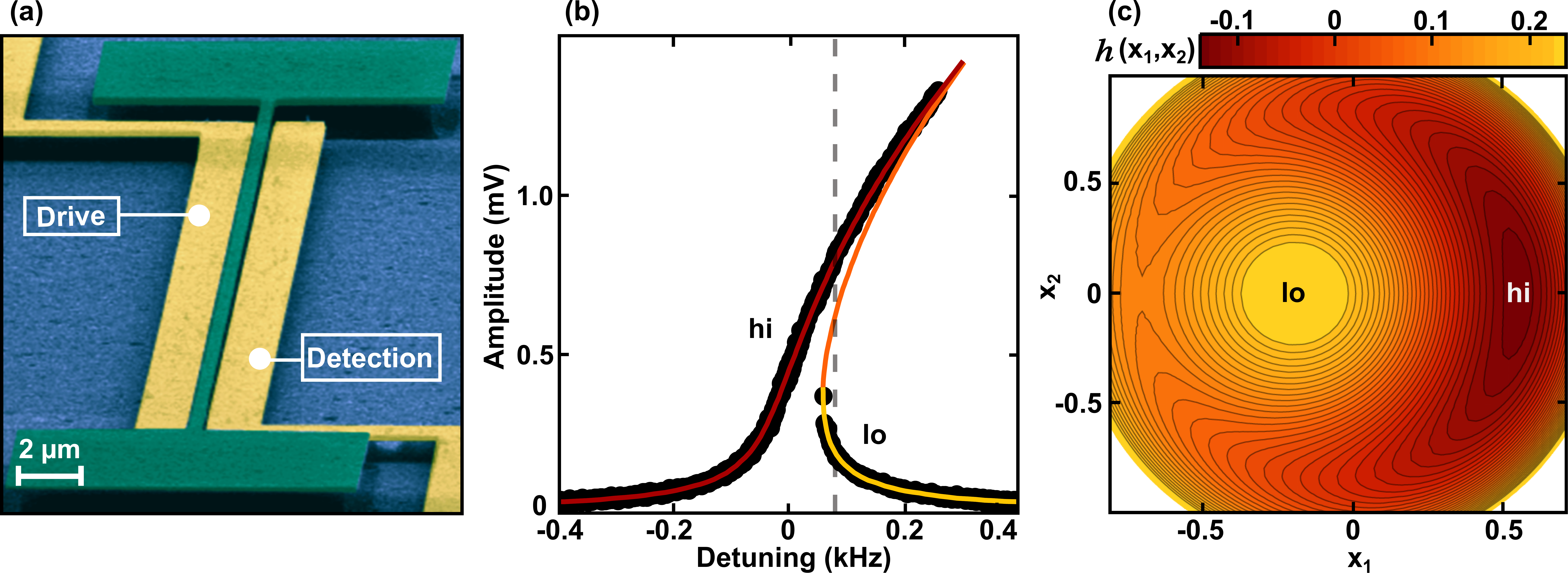}
\caption{ 
{\bf (a)}
Scanning electron micrograph of the doubly clamped silicon nitride string resonator (green) and two adjacent gold electrodes (yellow) for dielectric control. Schematic of electronic setup is detailed in the Sup. Mat. \cite{sup_mat}.
{\bf (b)}
Duffing response curve for an external drive of $-31$\,dBm (black dots) and 
fit of the Duffing model (Eq.~(\ref{eq:Duffing})).
The red (yellow) line denotes the high (low) amplitude solution marked ``hi'' (``lo''), 
while the orange line represents the unstable solution. Dashed gray line indicates the theoretically calculated critical switching point.
{\bf (c)} Phase space representation of the effective Hamiltonian function $h(x_{1},x_{2})$. Indicated are the high (``hi'', red) and low (``lo'', yellow) amplitude solution. 
The Hamiltonian function $h(x_{1},x_{2})$  scaled by $8\omega_{F}\delta\omega^2/(3\gamma)$ is plotted using the parameter  
$\sqrt3\gamma F^2/(32\omega_{F}^3\delta\omega^3)=0.013$, corresponding to the dashed line in (b), whereas the coordinate axes are scaled with $\sqrt{8\omega_{F}\delta\omega/(3\gamma)}$.
}
\label{fig:1}
\end{figure*}
%
%

%
%
%
%
\section{Experimental system}
The classical nanomechanical Duffing resonator is realized by a freely-suspended silicon nitride string fabricated on a fused silica substrate \cite{Faust:2012} 
whose material parameters are reported in literature \cite{Unterreithmeier2010,el-Hak2002}. 
The string under investigation is $270$\,nm wide, $100$\,nm thick and $55$\,\textmu m long, like the one depicted in Fig.~\ref{fig:1}a.
Owing to their strong intrinsic tensile pre-stress, these nanostring resonators exhibit ultra-high quality factors of several $100,000$ 
at room temperature \cite{Verbridge:2006ix,Faust:2012}. 
Dielectric transduction combined with a microwave cavity-enhanced heterodyne detection scheme is implemented via two adjacent gold electrodes also apparent in Fig.~\ref{fig:1}a.
The microwave cavity is pumped on resonance at $\approx 3.6$\,GHz to enable displacement detection while avoiding unwanted dynamical backaction effects.
The application of a dc voltage as well as a near-resonant rf drive tone $V_{\textrm{in}}$ 
enables actuation and eigenfrequency tuning of the string \cite{Unterreithmeier:2009gh,Faust:2012,Rieger2012}.
Moreover, the applied dc voltage also affects the damping rate and the nonlinearity, and introduces strong coupling between the out-of-plane and the in-plane mode when tuned into resonance. 
For all measurements discussed in the following, a constant dc voltage of $5$\,V is applied. 
Under that condition, the fundamental flexural out-of-plane mode can be considered independently, such that the following analysis is done in the single mechanical mode regime.       
The experiment is performed under vacuum at a pressure of $\leq10^{-4}$\,mbar and at room temperature of $293$\,K.

%
%
\section{Linear regime and characterization}
The sample is characterized by measuring response curves at various drive powers to calibrate the measurement, 
see also section II of the Supplementary Material (Sup. Mat.) \cite{sup_mat}.
A weak drive power allows for the characterization of the system in the linear regime.
The frequency response of the resonator is measured as a function of the frequency $f_F$ of the applied rf drive. 
The measured rf voltage signal is proportional to the resonator's amplitude. 
The resonance of the fundamental out-of-plane mechanical mode is found at $f_{0}=$\,6.529\,MHz with a linewidth of $2\Gamma/2\pi=20$\,Hz, 
yielding a quality factor of $Q \approx 325,000$. 
Note that this high quality factor is crucial for the presented work as it enables driving the resonator to amplitudes large enough to enter the nonlinear 
regime and to resolve the satellite peaks appearing in the power spectrum, as discussed in the following.

%
%
\section{Nonlinear regime and Duffing model}
Increasing the drive power leads to the well-known Duffing response \cite{Landau1986,Nayfeh:1995,Dykman:2012,Dykman:1984,Dykman1971}.
In this model the vibration of the single mode is described by the displacement $q(t)$ which obeys the equation
%
%
%
%
%
\begin{equation}
\label{eq:Duffing}
\ddot{q} + 2 \Gamma \dot{q} + \omega_0^2 q + \gamma q^3 = F \cos(\omega_F t) +\xi(t)
\, . 
\end{equation}
%
%
%
%
%
%
Here, $\omega_0=2\pi f_{0}$ is the angular eigenfrequency, $\Gamma$ the damping rate, $\gamma$ the nonlinearity parameter,
$F$ and $\omega_F = 2 \pi f_F$ are the amplitude and frequency of the external driving, and $\xi(t)$ is the thermal noise. 
The effective mass of the resonator is, for the time being, set to $m = 1$.
In a stationary vibrational state the coordinate $q(t) =A \cos\left( \omega_F t  + \theta \right)$
oscillates at the drive frequency with a phase $\theta$ with respect to the drive. The vibration amplitude $A$
is given by the solution of the cubic equation 
$A^2_{{j}}  \{ [\delta\omega - 3 \gamma A^2_{{j}} /(8 \omega_{0}) ]^2+ \Gamma^2  \}=F^2/ 4 \omega_{0}^2$, where $\delta\omega=\omega_F-\omega_0$ is the frequency detuning, $|\delta\omega|\ll \omega_0$ for the considered near-resonant driving.
The Duffing equation reflects the fact that the vibration frequency of a nonlinear resonator depends on its amplitude.
It can have one or three positive solutions. 
In the latter case, only the solutions with the largest and the smallest amplitude, $A_\mathrm{hi}$ and $A_\mathrm{lo}$, are stable.
An example of the measured amplitude as a function of the frequency detuning $\delta\omega$ is shown in Fig.~\ref{fig:1}b by black dots.
The solid line represents a fit of the Duffing model (Eq.~(\ref{eq:Duffing})). 

Only one fitting parameter, the Duffing nonlinearity parameter $\gamma$ is required, since the eigenfrequency $\omega_0$, the damping rate $\Gamma$, as well as the calibration of the driving force $F$ are known from the characterization in the linear regime \cite{sup_mat}. 
The nonlinear response curves obtained for different values of the rf drive power $P=-31$\,dBm (shown in Fig.~\ref{fig:1}b), 
$-30$\,dBm (shown in the Sup. Mat. \cite{sup_mat}), $-25$\,dBm, $-20$\,dBm and $-18$\,dBm are all fit using a single value of $\gamma$. 
As the amplitude of the resonator is measured in volts, the fit yields a nonlinearity parameter in units of $\mbox{V}^{-2} \mbox{s}^{-2}$. 
The obtained value, $9.28 \times 10^{16}\,\mbox{V}^{-2} \mbox{s}^{-2}$, can be converted into 
$\gamma = 1.54 \times 10^{26} \, \mbox{m}^{-2} \mbox{s}^{-2}$ using the amplitude conversion procedure described in the Sup. Mat. \cite{sup_mat}.

%
%
\section{Theory: squeezing in the power spectrum of a weakly damped oscillator}
The theoretical analysis of the resonator dynamics is done by  switching to the rotating frame,
$q(t) = \mathit{x}_1(t) \cos(\omega_F t)  +  \mathit{x}_2(t) \sin(\omega_F t)$ 
and
$\dot{q}(t) = - \omega_F \left[ \mathit{x}_1(t) \sin(\omega_F t)  -  \mathit{x}_2(t) \cos(\omega_F t) \right]$ 
where the quadratures $\mathit{x}_1(t)$ and $\mathit{x}_2(t)$ are new conjugate variables. 
Using the standard rotating wave approximation (RWA), 
one finds that the time evolution of these variables  is described by the equations 
%
%
%
%
%
\begin{align}
\dot{\mathit{x}}_1 &= \frac{\partial h \left(\mathit{x}_1,\mathit{x}_2\right) }{\partial \mathit{x}_2} - \Gamma  \, \mathit{x}_1  \, ,  \label{eq:Hamilton_equation_1}  \\
\qquad 
\dot{\mathit{x}}_2 &= - \frac{\partial h \left(\mathit{x}_1,\mathit{x}_2\right)}{\partial \mathit{x}_1} - \Gamma \, \mathit{x}_2   \, ,  \label{eq:Hamilton_equation_2}  
\end{align}
%
%
%
%
%
%
with the Hamiltonian function 
%
%
%
%
%
\begin{equation}
\label{eq:g_QP}
h \left(\mathit{x}_1,\mathit{x}_2\right) 
= 
\frac{3\gamma}{32\omega_F} {\left( \mathit{x}^2_1+ \mathit{x}^2_2 \right)}^2 
- \frac{\delta\omega}{2}  \left( \mathit{x}^2_1 + \mathit{x}^2_2 \right) - \frac{F}{2\omega_F} \mathit{x}_1 \, .
\end{equation}
%
%
%
%
%
In writing Eqs.~(\ref{eq:Hamilton_equation_1}), (\ref{eq:Hamilton_equation_2}) we have, for the time being, disregarded the noise. 
A contour plot of the function $ h \left(\mathit{x}_1,\mathit{x}_2\right) $  in the range of the bistability  is shown in Fig.~\ref{fig:1}c. 

A remarkable feature of our high Q nanostring resonator is that the damping rate $\Gamma$ is small not only compared to the eigenfrequency $\omega_0$, 
but also compared to the frequency detuning $\delta\omega$ and/or the typical frequency change due to the nonlinearity $\gamma A_{{j}}^2/\omega_F$. 
Therefore the damping can be treated as a small perturbation of the Hamiltonian dynamics of an auxiliary ``particle'' with coordinate $x_1$ and momentum $x_2$.
In this limit of weak damping, the extrema  ${j}=\mathrm{hi},\mathrm{lo}$ of $h$ correspond to the two stable states of forced vibrations \cite{Arnold:1989}.
At the extrema,  $x_{2,{{j}}}=0$, whereas $|x_{1,{{j}}}|=  A_{{j}}$ gives the vibration amplitude, if one disregards corrections $\propto \Gamma^2$.
The Hamiltonian dynamics for $\Gamma=0$ is characterized by the frequency $\omega_{{j}}$ of small-amplitude vibrations about the extrema of $h(x_1,x_2)$,
%
%
%
%
%
%
\begin{equation}
\label{eq:nu_i}
\omega_{{j}}^{\phantom{(1)}}
= 
\sqrt{ \omega_{{j}}^{(1)}\omega_{{j}}^{(2)}} \, , 
\end{equation}
%
%
where $\omega_{{j}}^{(1)}=3 \gamma A_{{j}}^2/8 \omega_F - \delta \omega$
and  $\omega_{{j}}^{(2)}=9 \gamma A_{{j}}^2/8 \omega_F  -\delta \omega$ 
(we note that $\omega_{{j}}^{(1,2)}$ can be positive or negative, but their product is positive). 
The frequency is different in the high- and low-amplitude states. 
In the considered weak-damping case $\Gamma \ll \omega_{{j}} $.

We now reintroduce noise into the equations for the quadratures 
and discuss thermal fluctuations about the stable states. 
 Even though the nanoresonator under investigation is small, thermal fluctuations at room temperature are weak. 
 If there is no driving ($F=0$ in Eq.~(\ref{eq:g_QP})), clearly $\langle  x_1 \rangle = \langle x_2 \rangle =0 $, while 
 the mean-square values of the quadratures are the same, 
 and for the considered weak nonlinearity $\langle  x_1^2\rangle = \langle x_2^2\rangle = k_BT/\omega_0^2$.

To analyze the squeezing of fluctuations about the states of forced vibrations for the case of weak damping, we linearize the equations of motion about the stable vibrational states 
$(x_{1{{j}}}, x_{2{{j}}})$ keeping the lowest-order terms in the decay rate $\Gamma$ 
(such linearization may be insufficient in the case of extremely weak damping, as discussed in Sec.~1. E of the Sup. Mat. \cite{sup_mat}). 
From Eqs.~(\ref{eq:Duffing}), (\ref{eq:Hamilton_equation_1}) and (\ref{eq:Hamilton_equation_2}), 
the resulting equations for the increments $\delta x_{1,2}$ in the presence of noise are, 
%
%
%
%
%
%
%
\begin{align}
\delta \dot{\mathit{x}}_1 &= \omega_{{j}}^{(1)} \delta\mathit{x}_2
- \Gamma \left( 1+\mu_{{j}} \right)  \delta\mathit{x}_1 + \xi_{x_1}(t)  \, ,
\label{eq:linear_1} \\ 
\delta \dot{\mathit{x}}_2 &= - \omega_{{j}}^{(2)}   \delta\mathit{x}_1
- \Gamma \left( 1 - \mu_{{j}} \right)  \delta\mathit{x}_2 + \xi_{x_2}(t)  \, .
\label{eq:linear_2}  
\end{align}
%
%
%
%
%
%
Here, $\mu_{{j}} = 6\gamma A_{{j}}^2/(3\gamma A_{{j}}^2-8\omega_F \,\delta\omega)$ and we have disregarded  terms $\propto \Gamma^2$.
Functions $\xi_{x_1}(t) $ and  $\xi_{x_2}(t) $ describe the noise that drives the quadratures. 
In the phenomenological model Eq.~(\ref{eq:Duffing}) these functions are given by the real and imaginary parts of $i\xi (t)\exp(i\omega_F t)/\omega_F$.
If the noise comes from the same coupling to a thermal  bath that leads to the vibration decay, on the time scale $\gg \omega_F^{-1}$  it is zero-mean, Gaussian 
and $\delta$-correlated, and the components $\xi_{x_1},\xi_{x_2}$ are independent and have equal intensity, $\langle \xi_{x_1}(t)\xi_{x_1}(0)\rangle = \langle \xi_{x_2}(t)\xi_{x_2}(0)\rangle = (2\Gamma k_BT/\omega_F^2)\delta(t-t')$. The power spectrum of the fluctuations of the oscillator coordinate in the approximation (\ref{eq:linear_1}) and (\ref{eq:linear_2}) is given by Eq.~(S13) of the Sup Mat \cite{sup_mat}.

A qualitative feature of the driven resonator is that the mean-square fluctuations of the in-phase and quadrature components of the coordinate are no longer equal and, for one of them, can be smaller than in the absence of the drive. This is the squeezing effect. In the considered case where the vibrations in the rotating frame are weakly damped, the mean square fluctuations in the state ${j} $ are (see the Sup. Mat. \cite{sup_mat})
%
%
%
%
\begin{align}
{\langle  \delta\mathit{x}^2_1 \rangle}_{ {j} } 
&= \frac{k_BT}{2m\omega^2_F}  \, \left( 1 + e^{-4\varphi_{{j}}} \right) \, , 
\label{eq:quadrature_1} \\
{\langle  \delta\mathit{x}^2_2 \rangle}_{ {j} }  
&=  \frac{k_BT}{2m\omega^2_F}  \, \left( 1 + e^{4\varphi_{{j}}} \right) \, , 
\label{eq:quadrature_2}
\end{align}
%
%
%
%
%
where the expression 
%
%
%
%
%
%
%
\begin{equation}
\exp(4 \varphi_{{j}})=\omega_{{j}}^{(2)}/\omega_{{j}}^{(1)}
\end{equation}
defines the squeezing parameter $\varphi_{{j}}$. 
Here, we have re-introduced the effective mass of the nanoresonator $m$ to facilitate the comparison with the experiment.
In the absence of driving, we find $A_{{j}}=0$ and thus $\varphi_{{j}}=0$, such that we recover the equipartition theorem, 
$\langle  \delta\mathit{x}^2_1 \rangle =\langle  \delta\mathit{x}^2_2 \rangle$.  
For the large-amplitude stable state $\varphi_{{j}} \equiv \varphi_{\mathrm hi} >0$, whereas for the small-amplitude state 
$\varphi_{{j}} \equiv \varphi_\mathrm{lo} <0$.
Obviously, the maximum squeezing attainable is a $50\,\%$ reduction of the squeezed quadrature according to 
Eqs.~(\ref{eq:quadrature_1}) and (\ref{eq:quadrature_2}).

Remarkably the squeezing appears directly in the power spectrum of the resonator \cite{Dykman2011,Dykman2012}. 
In the weak damping limit $\Gamma \ll  \omega_{{j}}$, one obtains 
%
%
%
%
%
%
%
%
\begin{align}
Q_{{j}}(\omega) 
&\approx 
\frac{\Gamma k_BT}{4\pi m \omega_F^2}  \,\,
\frac{\cosh 2\varphi_{{j}}
(\cosh 2\varphi_{{j}} \pm 1)}{(\omega-\omega_F \mp{\cal S}_{{j}}  \omega_{{j}})^2 + \Gamma^2}  \nonumber \\
\mbox{for}& \quad  \left|\omega-\omega_F \mp {\cal S}_{{j}} \omega_{{j}} \right| \ll  \omega_{{j}} \, ,
\label{eq:S_w_app_1} 
\end{align}
with ${\cal S}_\mathrm{hi}=+1$ for the large-amplitude stable state and ${\cal S}_\mathrm{lo}=-1$ for the small-amplitude stable state, respectively ~\cite{sup_mat}. 
The power spectrum $Q_{{j}}(\omega)$ consists of two Lorentzian peaks centered at the frequencies 
$\omega_F \pm {\cal S}_{{j}} \omega_{{j}}$ 
with the half width given by the damping rate of resonator in the absence of driving $\Gamma$.
They can be thought of as the Stokes and anti-Stokes components of the Raman scattering of the driving field by the small-amplitude vibrations 
of the resonator near the corresponding stable state.  
Importantly, the very state is formed by the drive. 
The ratio of the intensities of the satellite peaks 
%
%
%
%
%
\begin{align}
\mathcal{I}^{(+)}_\mathrm{hi} / \mathcal{I}^{(-)}_\mathrm{hi}   & = 1/\tanh^2 \left(  \varphi_\mathrm{hi} \right)  \label{eq:ratio_area_1}  \, , \\
\mathcal{I}^{(+)}_\mathrm{lo} /  \mathcal{I}^{(-)}_\mathrm{lo}  & = \tanh^2 \left(  \varphi_\mathrm{lo} \right) \, ,   \label{eq:ratio_area_2}
\end{align}
is determined by the squeezing parameter $\varphi_{{j}}$. The squeezing parameter can thus be direcly found from the power spectrum. An advantageous feature of the ratios Eqs.~(\ref{eq:ratio_area_1}), (\ref{eq:ratio_area_2}) is their independence of the temperature. 
Therefore even if the nanoresonator is slightly heated by the drive, they should not change.

We emphasize that the peak intensities ${\cal I}^{(\pm)}$ are well defined if the satellite peaks are well resolved. This condition is met, as seen from Eq.~(\ref{eq:S_w_app_1}), provided the widths of the peaks are small compared with the distance between them, i.e., $\Gamma\ll \omega_{{j}}$. The latter inequality has a simple physical meaning: in the rotating frame, the vibrations about the stable state of forced oscillations of the nonlinear resonator are underdamped, see also Sec. I.D and I.E of the Sup. Mat. \cite{sup_mat}. This is a stronger condition than the condition that the nanoresonator mode is underdamped in the laboratory frame, i.e., $\Gamma\ll \omega_0$. However, for weakly damped nonlinear nanoresonators of current interest, including the one studied in this paper, the condition $\Gamma\ll  \omega_{{j}}$ holds in a broad range of the amplitudes and frequencies of the driving field.  

The relations (\ref{eq:ratio_area_1}) and (\ref{eq:ratio_area_2}) do not hold in the quantum regime, $\hbar\omega_0\gtrsim k_BT$. The power spectrum (i.e., the fluorescence spectrum) is symmetric with respect to the drive frequency for $k_BT\ll \hbar\omega_0$ \cite{Drummond:1980c}. However, quantum fluctuations of the driven nonlinear mode are squeezed. The variances of the in-phase and quadrature components are different. In the strongly underdamped regime discussed here, there is an alternative spectral measurement that allows one to find the squeezing parameter both in the classical and quantum regimes. This measurement involves driving the mode by an additional weak probe drive $F'\exp(-i\omega' t)$ at frequency $\omega'$ close to the strong-drive frequency $\omega_F$. Such drive leads to an additional term in the mode displacement, which oscillates at frequencies $\omega'$ and $2\omega_F-\omega'$,  $\delta \langle q(t)\rangle = \chi(\omega')F'\exp(-i\omega' t) + {\cal X}(\omega')F'\exp[-i(2\omega_F-\omega')t]$ \cite{Dykman:1979,Dykman:1994gl}. In a nanomechanical resonator, spectral peaks at the frequency of the probe drive  have been observed in Ref.~\onlinecite{Antoni2012}. The susceptibility $\chi(\omega)$ directly reveals the squeezing in the strongly-underdamped regime. Both Im~$\chi(\omega)$ and $|\chi(\omega)|^2$ display two narrow sideband peaks, with the ratio of their areas determined by the squeezing parameter \cite{sup_mat}. Squeezing of quantum fluctuations about a metastable state occurs also in a driven oscillator resonantly coupled to a two-level system \cite{Peano2010a}.

%
%
\section{Experimental observation of the thermal squeezing in the power spectrum}
\begin{figure}[t!]
	\includegraphics[scale=0.27]{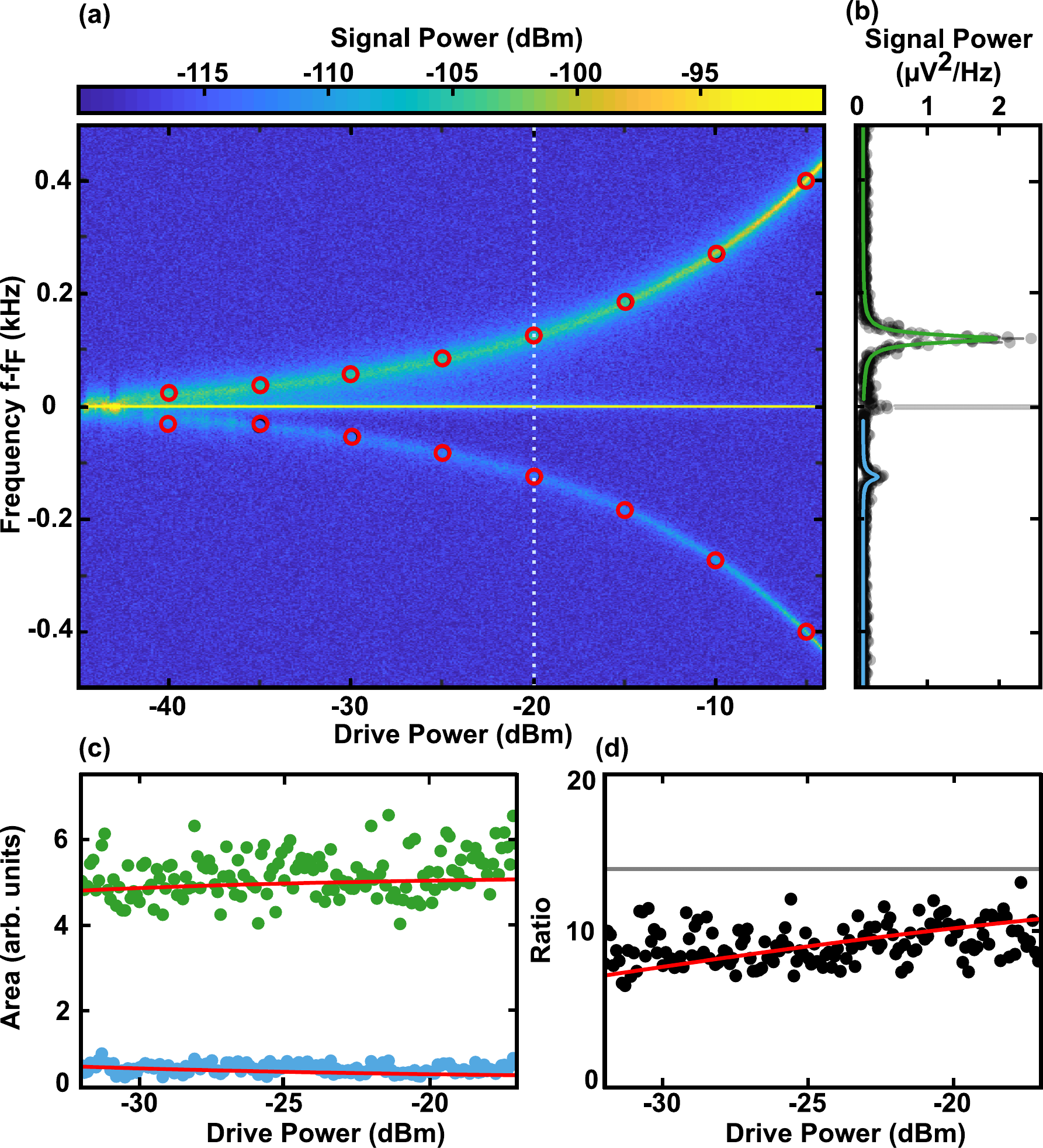}
	\caption{
		{\bf (a)} 
		Color coded power spectra showing the increasing splitting of the satellite peaks with the increasing drive power for the drive frequency $f_F=f_0$ where the resonator is monostable. 
		Red open circles denote the calculated positions of the satellite peaks. The central line at  $f-f_F=0$ is plotted with a reduced brightness to improve the visibility of the satellites.
		{\bf (b)} 
		Linecut along the white dotted line  in Fig.~\ref{fig:2}a 
		illustrating the satellite peaks as well as their Lorentzian fits for a drive of $-20$\,dBm. 
		The central line at $f=f_F$ (gray line) is truncated.
		{\bf(c)} Area of the high (green) and low (blue) frequency satellite peaks extracted from the Lorentzian fits as a function of the drive power. 
		Red lines show the theoretical prediction that takes into account the partial overlap of the peaks.  
		{\bf (d)} Ratio of the  areas of the satellite peaks as a function of drive power. 
		Red and gray lines show, respectively, the theoretical prediction that takes into account the partial overlap of the peaks 
		(see Sup. Mat.  \cite{sup_mat})
		and the one based on Eq.~(\ref{eq:ratio_area_1}).
	}
	\label{fig:2}
\end{figure}

%
%
%
%
To validate these theoretical findings, we apply a resonant sinusoidal drive tone to the fundamental flexural mode of the nanostring ($f_{F}=f_{0}$) and record power spectra for different drive powers using a spectrum analyzer operated in the FFT-mode. 
Under resonant driving, the resonator has one stable vibrational state, with the parameters in 
Eqs.~(\ref{eq:nu_i}-\ref{eq:S_w_app_1})
corresponding to the high-amplitude state $A_\mathrm{hi}$.
Figure~\ref{fig:2}a displays power spectra for drive powers in the range between $-45$\,dBm and $-5$\,dBm, with a color coded signal power (dBm). 
The bright, narrow line centered at zero corresponds to forced vibrations at $f_F$. 
The drive tone is flanked by two satellite peaks. Their separation from the drive tone is symmetric and increases with drive power.
We identify these sideband-resolved satellite peaks with the  thermal noise-induced small-amplitude vibrations around the stable state of the driven resonator. 
Thus the peaks should be centered at the frequencies  $\omega_F \pm \omega_{\mathrm{hi}}$. 

%
%
%
%
%
%

The experimentally observed satellite peaks are compared with the theoretical prediction of Eq.~(\ref{eq:nu_i}) in Fig.~\ref{fig:2}a, where the calculated positions of the peaks  are shown as open red circles. 
For better visualization, only a few distinct points are plotted. 
We find the experimental data to coincide with the theory, and also recover the expected scaling of the splitting of the satellite peaks with the drive power $\omega_{\mathrm{hi}}   \propto  A_{\mathrm{hi}}^2 \propto F^{2/3} \propto P^{1/3}$. 

Another remarkable feature of the satellite peaks is apparent from their intensities.
Figure~\ref{fig:2}b depicts a line cut extracted from Fig.~\ref{fig:2}a 
at $-20$\,dBm. 
Each peak is fitted by a Lorentzian with a linewidth of $2\Gamma/2\pi=20$\,Hz, as shown in Fig.~\ref{fig:2}b. 
As predicted by the theoretical model, this linewidth coincides with that of the linear resonance of the string \cite{sup_mat}.
Clearly, the satellite peak at higher frequency is much brighter than that at the lower frequency. 
This observation is in agreement with the theoretical model, which predicts non-equal  intensities of the satellite peaks 
as a result of the classical squeezing of thermal fluctuations. 

More precisely, as outlined in Eq.~(\ref{eq:ratio_area_1}) for the high-amplitude state $A_\mathrm{hi}$, 
a higher intensity is expected for the  satellite peak at the higher frequency. 
Following the model, the ratio of the areas enclosed by the peaks is simply related to the squeezing parameter
$\varphi_{{j}}$. 
The  areas extracted from the fit are plotted in Fig.~\ref{fig:2}c 
as a function of the drive power, where green corresponds to the brighter, higher frequency peak and blue to the lower frequency peak. 
The experimental data are compared with the theoretical predictions which are shown in Fig.~\ref{fig:2}c by the red lines \cite{sup_mat}.
As suggested by the theoretical model, a pronounced difference in the areas is observed.
The ratio of the areas is plotted in Fig.~\ref{fig:2}d, and again, we find very good agreement between the experimental data (black dots) and the theoretical predictions (red line).

The theoretical calculations of the areas and their ratio shown in Fig.~\ref{fig:2} are 
obtained from a more general analysis of the power spectrum.
This analysis is not limited to the condition $\Gamma\ll \omega_j$ and thus takes into account the overlapping of the satellite peaks. 
It is provided in Secs.~I~C and I~D of the Sup. Mat.  \cite{sup_mat}. The ratio of the areas for the limit of small damping, Eqs.~(\ref{eq:ratio_area_1}) and (\ref{eq:ratio_area_2}), is also included in Fig.~\ref{fig:2}d as a gray line. 
In this limit the ratio
is independent of the drive power; it provides the fundamental limiting value for the ratio of the areas of the satellite peaks.
For our high-Q nanostring resonator, the measured ratio approaches this value as the separation of the peaks increases with the increasing drive power.

\begin{figure}[t!]
	\includegraphics[scale=0.24]{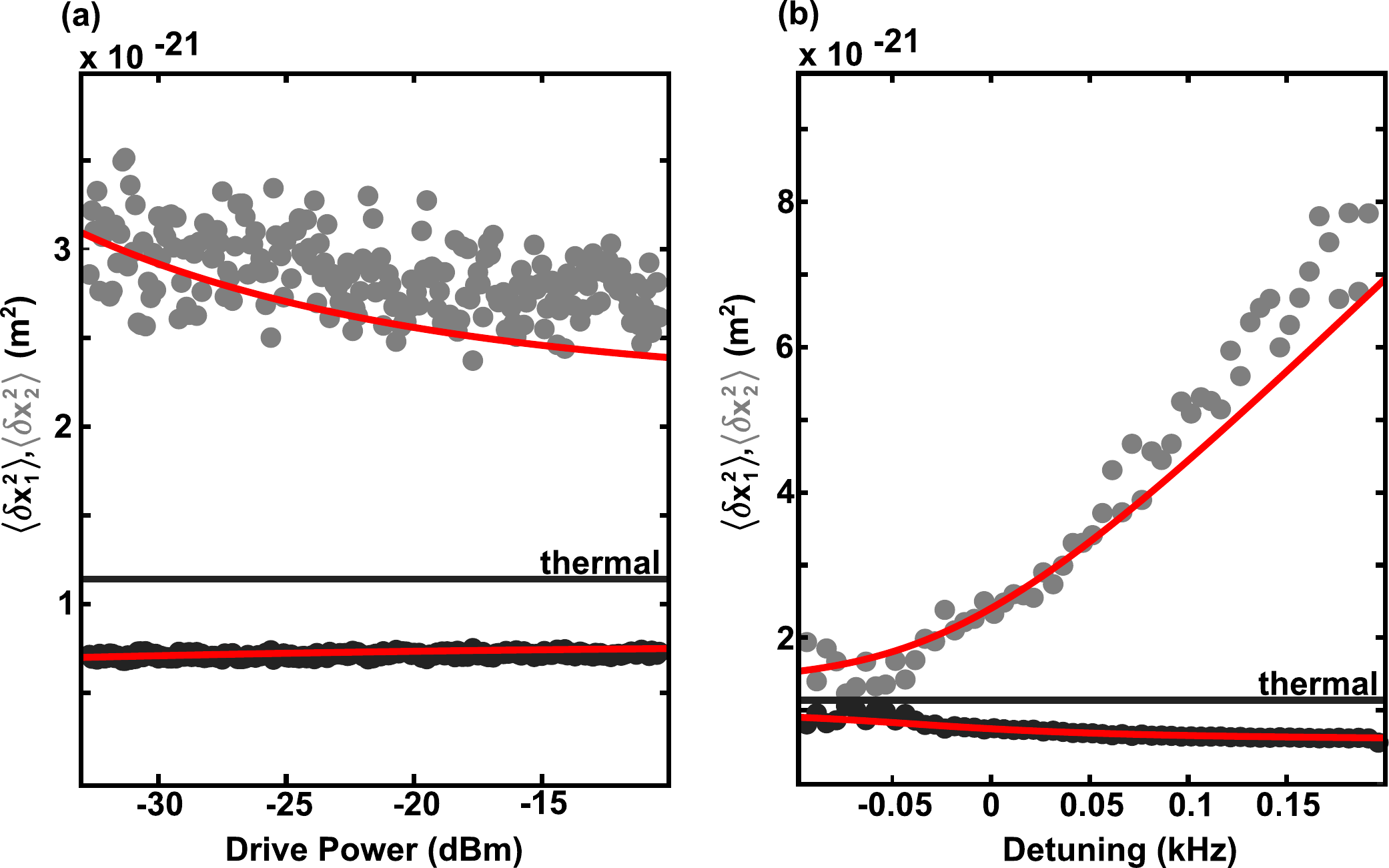}
	\caption{
		Variance of in-phase and quadrature fluctuations around the stable state of forced vibrations as a function of 	{\bf (a)} drive power and 	{\bf (b)} detuning. Black and gray dots show the in-phase and quadrature values extracted from the experimentally determined satellite area ratio, respectively, whereas red lines show the corresponding theoretical model including the partial overlap of the peaks (no free parameters, see Sup. Mat. \cite{sup_mat}). Black lines in (a) and (b) indicate the thermomechanical fluctuations at $293$\,K. }
	\label{fig:quadratures}
\end{figure}

The squeezing parameter $\varphi_\mathrm{hi}$ extracted from the areas of the satellites discussed in Fig.~\ref{fig:2}c,d can be employed to compute the mean square fluctuations of the in-phase and quadrature component of the stable state of forced vibrations using Eqs.~(\ref{eq:quadrature_1}) and (\ref{eq:quadrature_2}). Figure~\ref{fig:quadratures}a compares the experimentally obtained fluctuations ($\langle  \delta\mathit{x}^2_1 \rangle_{ \mathrm{hi}} $ and $\langle  \delta\mathit{x}^2_2 \rangle_{ \mathrm{hi}}  $  represented as black (gray) dots, respectively) with the theoretical model accounting for the partial overlap of the satellite peaks \cite{sup_mat} (red lines). The mean square of the thermomechanical fluctuations at $293$\,K is included as a black solid line, clearly showing that a significant squeezing of the in-phase quadrature is accomplished.

%
%
%
%
%
%

%
%
%
%
According to the theory, the satellite peaks in the power spectrum also depend on the detuning of the drive frequency $f_{F}-f_{0}$. 
We therefore repeat the measurement routine, now for a fixed drive power of $-20$\,dBm and a variable detuning of the drive. 
The resonator is initialized in the high-amplitude state by sweeping up the drive frequency from $30$\,kHz below $f_0$ to the desired $f_F$ before recording the power spectrum.

\begin{figure*}[t!]
\includegraphics[scale=0.29]{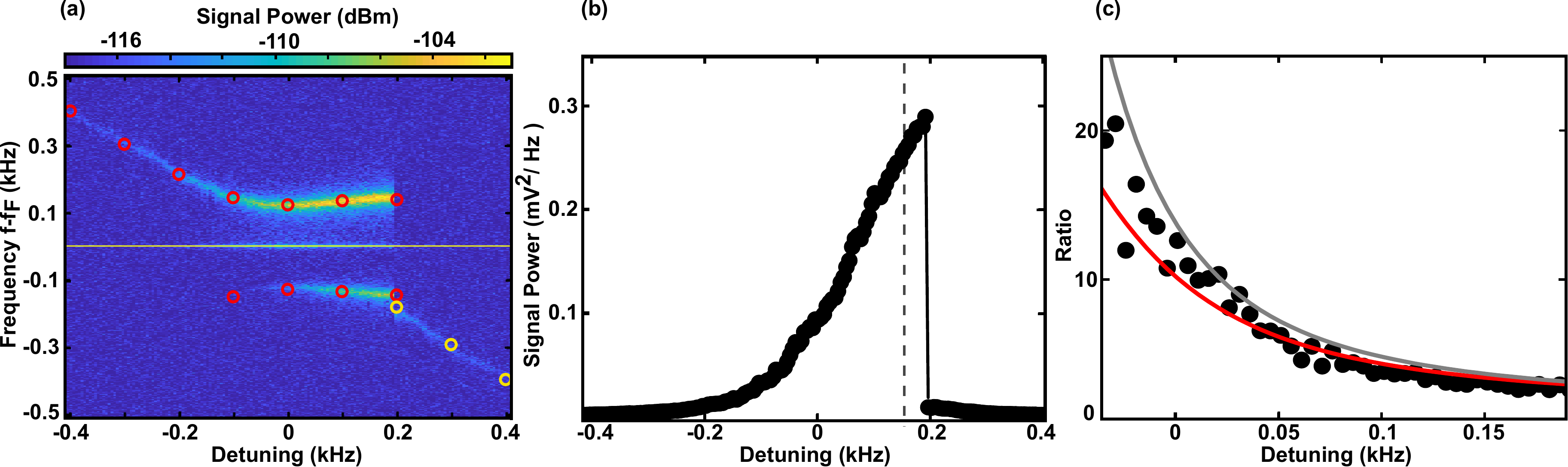}
\caption{
{\bf (a)} Color coded power spectra showing the positions and intensities of the satellite peaks as a function of the detuning of the drive frequency $f_F-f_0$ for the drive power -20~dBm. The central line at  $f-f_F=0$ is plotted with a reduced brightness to improve the visibility of the satellites.
Open circles denote the calculated  positions for the high (red) and low (yellow) amplitude states (see Fig.~\ref{fig:1}b).
{\bf (b)} The power of the signal at the drive frequency $f_F$ as a function of the detuning. The discontinuity observed at a detuning of $190$\,Hz indicates the switching of the resonator from the high- to the low-amplitude state for a slow ramp up of the detuning. It coincides with the discontinuity of the satellite peaks in Fig.~\ref{fig:3}a.
The dashed gray line indicates the theoretically calculated critical switching point.
{\bf (c)} Ratio of the  areas of the satellite peaks  for the high-amplitude state as a function of the detuning.
Red and gray lines show the theoretical predictions with and without the 
small overlapping of the satellite peaks, respectively.
}
\label{fig:3}
\end{figure*}

Figure~\ref{fig:3}a displays the power spectra as a function of the detuning $f_{F}-f_{0}$.
For large negative detuning, $f_F-f_0<0$, only the satellite peak at a higher frequency can be discerned; its distance from the drive tone $f_F$ increases with the increasing $-(f_F-f_0)$.
For small detuning, both satellite peaks are resolved. 
They are at equal distances from $f_F$, which only slightly increase 
with $f_F-f_0$ for $f_F-f_0>0$. In contrast, the intensities of the peaks are increasing.
The splitting at zero detuning equals the one shown in the resonantly driven case discussed in Fig.~\ref{fig:2}a for a drive power of $-20$\,dBm. 

%
%
%
%
%
%
%
%

Interestingly, the satellite peak at higher frequency  vanishes abruptly for the detuning of $190$\,Hz, whereas the lower frequency one remains. However, the  lower frequency peak exhibits a discontinuity at $190$\,Hz, and continues with a larger splitting, a different slope, and a strongly reduced intensity. At the same detuning of $190$\,Hz the amplitude at the drive tone drops to a drastically smaller value, as shown in Fig.~\ref{fig:3}b. This is a signature of the resonator switching from the high-amplitude state $A_{\textrm{hi}}$ to the low-amplitude state  $A_{\textrm{lo}}$.  The displayed signal power has been extracted from a linecut in Fig.~\ref{fig:3}a at the driving frequency, $f=f_F$. Since the measurement routine to record each of the power spectra in Fig.~\ref{fig:3}a exposes the resonator to the drive for more than one minute, this represents a much slower measurement than a typical (Duffing) response curve measurement such as the one shown in Fig.~\ref{fig:1}b. 

The observed satellite peaks on the opposite sides of the critical detuning $\Delta f_{\rm cr}\equiv (f_F-f_0)_{\rm cr}\approx 190$~Hz are associated with the high- and low-amplitude state $A_{\textrm{hi}}$ and $A_{\textrm{lo}}$ of the resonator. 
They are compared in Fig.~\ref{fig:3}a with the theoretical prediction for the two stable states, 
which are superposed on the measured data as red and yellow open circles, respectively. 
In the both states, we find the experiment and the theory to coincide completely.

We repeat the analysis described for the resonantly driven case of Fig.~\ref{fig:2} 
and extract the areas of the high and low frequency satellite peaks for each power spectrum by fitting two Lorentzians (not shown). 
When the resonator is in the high-amplitude state, i.e. for a detuning below $\Delta f_{\rm cr}$, both satellite peaks are resolved and appear for a certain range of detunings. The ratio of the obtained  areas of the  peaks for this detuning is shown in Fig.~\ref{fig:3}c as black dots. Like for the resonantly driven case, this quantity is associated with the squeezing parameter. 

According to the theory of Sec.~V, the ratio of the areas of the peaks depends on the detuning of the drive frequency. 
For the high-amplitude stable state, it is asymmetric with respect to $f_0$ and decreases as the detuning varies from negative to positive.
The experimental data in Fig.~\ref{fig:3}c is compared with the theoretical prediction for the weak-damping limit, Eq.~(\ref{eq:ratio_area_1}) (gray line), 
and for the more general approximation that takes into account the small overlapping of the satellite peaks \cite{sup_mat} (red line). 
Once more, the agreement between the experiment and the theory is remarkable.
The resulting mean square fluctuations of the in-phase and quadrature component about the stable state of forced vibrations are presented in Fig.~\ref{fig:quadratures}b, again demonstrating the squeezing of the in-phase quadrature with respect to the thermomechanical fluctuations.

Above the switching point, $f_F - f_0>\Delta f_{\rm cr}$ the resonator is in the low-amplitude state, 
and only one satellite peak is resolved. 
Therefore the  ratio of the areas of the peaks and thus the squeezing parameter cannot be evaluated here.
Notice, however, that the data clearly shows the anticipated reversal of the  intensities of the satellite peaks between the two stable solutions, as predicted 
by Eq.~(\ref{eq:ratio_area_1}) and (\ref{eq:ratio_area_2}): 
While the high frequency satellite peak has a higher intensity for the high-amplitude stable state, the low frequency peak  
is the dominating one for the low-amplitude state. 
In addition, while the ratio of the areas of the peaks for the high-amplitude state has decreased to a value $\approx 1$ in the vicinity of the switching point $f_{\rm cr}$, for the low-amplitude 
state the ratio is large, according to the theory, which explains why the low frequency satellite peak  is resolved whereas the high frequency peak cannot be detected. 

For a positive or negative detuning exceeding $400$\,Hz, Fig.~\ref{fig:3}a 
exhibits only one peak, and the slope of its frequency vs the drive frequency is $-1$. Such slope and a single peak in the power spectrum are expected for an oscillator in the absence of a driving force. 
Experimentally, for still larger detuning, we are not able to resolve thermal motion of the driven resonator, as is also the case for the undriven resonator. We attribute this to an insufficient displacement sensitivity of the detection 
setup far away from the driving frequency or in the absence of the drive, 
while the thermally-induced spectral features are resolved near $f_F$. 
Apparently, the displacement sensitivity increases in the presence of the driving, 
which is likely a consequence of our heterodyne microwave-cavity assisted displacement detection scheme~\cite{Faust:2012}.

%
%
\section{Critical switching point}
Finally, we discuss the switching between the two stable states of the Duffing resonator. 
It is characterized by two rates, that from the high-amplitude to the low-amplitude state, $W_{\textrm{hi}\rightarrow\textrm{lo}}$, 
and that from the low-amplitude to the high-amplitude state, $W_{\textrm{lo}\rightarrow\textrm{hi}}$.
At the critical frequency detuning these rates 
are equal, $W_{\textrm{hi}\rightarrow\textrm{lo}}=W_{\textrm{lo}\rightarrow\textrm{hi}}$. Respectively, the stationary populations of the stable states are also equal. The rates change with the parameters exponentially strongly. Therefore, away from the critical value of the detuning, the populations of the states are strongly different and only one state is ``visible''. If the detuning is slowly varied across the critical value, the oscillator should switch from one state to the other in a very narrow range. 
For weak damping, $\Gamma \ll \delta\omega$, the theoretical value of the critical detuning~\cite{Dykman:1979}, in terms of the parameters of the studied nanoresonator, is 
$\Delta f_{\rm cr}  \approx  904.6 {\rm s}^{-1} (V_{\textrm{in}}[V])^{2/3}$.
It is shown in Fig.~\ref{fig:1}b as a vertical dashed gray line at $V_{\textrm{in}}=17.8$\,mV ($-31$\,dBm).

Experimentally, the interchange of the most probable states at the critical point can only be observed in a slow measurement.
Clearly, the response curve shown in Fig.~\ref{fig:1}b 
does not reveal this point, since the detuning was swept in the both directions fast enough to allow the system to stay in the metastable 
high- or low-amplitude state well beyond the critical point, until close to the bifurcation point.

In contrast, the response curve shown in Fig.~\ref{fig:3}b 
results from a much slower measurement as described above. This allows the resonator to approach its most probable stable state for every applied detuning and clearly demonstrates sharp switching between occupying practically one or the other state. 

The switching point observed in Fig.~\ref{fig:3}b is expected to be close to the theoretical critical switching point, which is shown by a dashed gray line ($ V_{\textrm{in}}=65$\,mV).
Indeed, the difference between the experimental and theoretical values is only $40$\,Hz. 
This difference can be attributed to a slight non-adiabaticity of the frequency sweep.  Futhermore, given the statistical nature of the switching, slow room temperature fluctuations cannot be ruled out as an alternative source of the discrepancy, because the effect of a typical eigenfrequency drift of almost $1$\,kHz$/$K could not be completely eliminated, even though the eigenfrequency was re-determined prior to every measurement. In the future, the results can be extended to measure the individual switching rates using different sweep times \cite{Aldridge2005,Stambaugh2006,Defoort2015}.

%

We emphasize that, in the regime we have studied, the driven resonator has no detailed balance. 
Understanding  fluctuation-induced transitions between the stable states of  systems lacking detailed balance, i.e., generically, for all systems away from thermal equilibrium, is of interest for various areas of physics, chemistry, and biology. 
The weak-damping regime attained in the present work is particularly important, as the phase space of the system is two-dimensional rather than the effectively one-dimensional phase space close to bifurcation points. A high-dimensional phase space significantly complicates the theoretical analysis of the switching rate. 
To the best of our knowledge, the present results show the first quantitative comparison with analytical results obtained for systems lacking detailed balance.

%
%
\section{Conclusions}
In conclusion, we report a new manifestation of squeezing of thermal fluctuations in a broad parameter range of a resonantly driven nanomechanical mode. The squeezing is indirectly determined by measuring the power spectrum of the mode in the sideband-resolved regime, where the spectrum exhibits two well-separated peaks symmetrically positioned with respect to the drive frequency. 
The peaks can be thought of as Stokes and anti-Stokes component in a Raman scattering picture with the caveat that the underlying process is multi-photon, as multiple photons of the resonant driving field are involved. 

The sidebands feature unequal intensities. The ratio of the intensities is determined by the squeezing parameter.
It was directly read out from the experimental data, thus providing a novel way not only to infer, but also to quantitatively characterize squeezing.

Our findings are supported by a theoretical model which is in excellent agreement with the experimental data with no free parameters.
The model shows that, for the resonantly driven underdamped Duffing resonator, the squeezed quadrature can be suppressed by a factor of 2, giving rise to a $3$\,dB 
limit, as in the case of parametrically induced squeezing \cite{Rugar:1991he}.
Importantly, no fine tuning to a specific operation point is required for obtaining squeezing in a high-quality-factor resonator.

Squeezing of thermal fluctuations about the state of forced vibrations in weakly damped nonlinear systems is a generic concept
as it is related to the breaking of the continuous time translation symmetry by the drive. 
The same applies to the asymmetry of the power spectrum and the response spectrum. Therefore, the squeezing and the asymmetry are intrinsically related to each other and we use one of them to characterize the other.

At the same time, it should be noted that the spectral characterization 
of the squeezing is an indirect one. For applications in precision sensing, care should be taken to ensure that no extra noise is added by the measurement setup.

An important advantageous feature of characterizing squeezing of thermal (and quantum) fluctuations in driven mesoscopic vibrational systems from a spectral measurement is its insensitivity to weak frequency noise. This is important both for nanomechanical resonators as the ones studied here, and also for microwave cavity modes. In these systems, the mode eigenfrequencies display slow fluctuations with $1/f$-type spectrum. Such fluctuations lead to a small broadening of the spectral peaks and a very small change of the peak intensities. Thus they make a small effect on the measured squeezing parameter. In contrast, they significantly complicate the homodyne measurement for weakly damped systems, as discussed in Sec.~2.F of the Sup. Mat. \cite{sup_mat}.

A promising application is the possibility of employing driven weakly damped modes as detectors of weak signals at frequency 
$f_S$ close to the drive frequency, $|f_F-f_S| \approx \omega_{j}$. Driven modes can resonantly amplify such signals,
which can be thought of as a multi-photon analog of stimulated Raman scattering. The amplification is determined by the squeezing \cite{sup_mat}, which in turn allows one to determine the squeezing parameter from the response spectrum. 
Importantly, the corresponding spectral measurement can be done also in the quantum regime \cite{sup_mat}, where, as shown in Ref.~\onlinecite{Drummond:1980c}, the sidebands in the  emission spectrum  are symmetric independent of the squeezing parameter.

\section{Acknowledgements}
Financial support by the Deutsche Forschungsgemeinschaft via the collaborative research
center SFB 767, the European Union’s Horizon 2020 Research and Innovation Programme under Grant
Agreement No 732894 (FET Proactive HOT), and the German Federal Ministry of Education and Research
(contract no. 13N14777) within the European QuantERA cofund project QuaSeRT is gratefully acknowledged.
M.~I.~D. also acknowledges support from the Zukunftskolleg
Senior Fellowship at the University of Konstanz and from the National Science Foundation (Grant No. DMR-1806473).


%

\end{document}



\title{Supplemental Material:\\ Spectral evidence of squeezing of a weakly damped driven nanomechanical mode}

\author{J.\, S. Huber }
\affiliation{Department of Physics, University of Konstanz, 78457 Konstanz, Germany}
\author{G. Rastelli}
\affiliation{Department of Physics, University of Konstanz, 78457 Konstanz, Germany}
\author{M.\, J. Seitner}
\affiliation{Department of Physics, University of Konstanz, 78457 Konstanz, Germany}
\author{J.\, K\"olbl}
\altaffiliation{present address: Department of Physics, University of Basel, 4056 Basel, Switzerland}
\affiliation{Department of Physics, University of Konstanz, 78457 Konstanz, Germany}
\author{W. Belzig}
\affiliation{Department of Physics, University of Konstanz, 78457 Konstanz, Germany}
\author{M.\,I. Dykman}
\altaffiliation{dykmanm@msu.edu}
\affiliation{Michigan State University, East Lansing, Michigan 48824, USA}

\author{E.\,M. Weig}
\altaffiliation{eva.weig@uni-konstanz.de}
\affiliation{Department of Physics, University of Konstanz, 78457 Konstanz, Germany}

\begin{abstract}
In this Supplemental Material we provide a theoretical analysis of the thermal noise squeezing and how such squeezing shows in the power spectrum 
of a driven Duffing resonator, as well as the manifestation of squeezing in the spectra of response to an extra weak field, which is pronounced in the both classical and quantum regimes. 
We also discuss the characterization of the nanomechanical resonator, the calibration of the detection setup, the analysis of the data provided 
in the main text, and present further supplementary experimental data. We conclude by demonstrating the impact of frequency fluctuations on the direct homodyne detection of the noise quadratures of a weakly damped nanomechanical resonator.
\end{abstract}

\date{\today}

\maketitle
%
%
%
\section{Theory}
%
%
%
%
%
%
%
\subsection{The Duffing model}
The fundamental mode of the silicon nitride nanostring resonator under investigation is well described by the model of the Duffing resonator. 
The beam displacement $q$ at the antinode follows the equation
%
%
%
%
%
\begin{equation}
\label{eq:Duffing}
\ddot{q} + 2 \Gamma \dot{q} + \omega_0^2 q + \gamma q^3 = F \cos(\omega_F t)  + \xi(t)  \, , 
\end{equation}
%
%
%
%
in which $\Gamma$ is the damping coefficient, $\omega_0$ is the mode eigenfrequency, $\gamma$ is the Duffing nonlinearity parameter, $F$ is the amplitude of the external driving, and $\omega_F$ is the driving frequency.  
In our experiment we have $\gamma>0$. The vibrations have inversion symmetry, and therefore there is no term $\propto q^2$ in Eq.~(\ref{eq:Duffing}). 
Here for brevity we have set the effective mass of the resonator at the position of the antinode $m=1$; 
alternatively, one can think that the forces $F$ and $\xi(t)$ incorporate the factor $1/m$.

The term $\xi(t)$ represents the thermal noise. 
Since the frequency $\omega_0$ is small compared to $k_BT/\hbar$ and compared to the reciprocal correlation time 
of the thermal reservoir (thermal phonons, in our system), the dissipation has no delay and the noise $\xi(t)$ is zero-mean, 
Gaussian, and $\delta$-correlated. 
With the effective mass taken explicitly into account, the noise correlator reads  $\langle  \xi(t)  \rangle =0 $ 
and $\langle  \xi(t) \xi(t') \rangle = \delta(t-t') \, 4 \Gamma k_B T / m $, where $T$ is the room temperature.

It is convenient to switch to the rotating frame and to introduce the scaled complex vibration amplitude $\mathit{y}$, 

%
%
%
%
\begin{equation}
\label{eq:q_gamma}
q(t)=\sqrt{ \frac{2\omega_F \Gamma}{3 \gamma} } 
\left[  \mathit{y}(t) e^{i\omega_F t} +  \mathit{y}^*(t)e^{-i\omega_F t} \right] \, , \quad
\dot{q}(t)=\sqrt{ \frac{2\omega_F \Gamma}{3 \gamma} }  (i \omega_F) 
\left[  \mathit{y}(t) e^{i\omega_F t} -  \mathit{y}^*(t)e^{-i\omega_F t} \right] \, .
\end{equation}
The relations with the two quadratures defined in the main text are simply 
$\mathit{x}_1 = \sqrt{8\omega_F \Gamma/(3\gamma)} \,\mbox{Re} \, \mathit{y}$ and 
$\mathit{x}_2 = -  \sqrt{8\omega_F \Gamma/(3\gamma)} \,\mbox{Im} \, \mathit{y}$.
In the rotating wave approximation (RWA), we obtain from Eq.~(\ref{eq:Duffing}) the equation of motion for $\mathit{y}(t)$ in the form
%
%
%
%
\begin{equation}
\label{eq:Duffing-RWA-u}
\dot{\mathit{y}} = i\frac{\partial h_{\mathit{y}}}{\partial \mathit{y}^*}  -\Gamma \mathit{y}  
-i\Gamma\xi_{\mathit{y}} (t); \qquad h_{\mathit{y}}=\Gamma\left[\frac{1}{2}|\mathit{y}|^4 - \Omega |\mathit{y}|^2 - \beta_{\mathit{y}}^{1/2}(\mathit{y}+\mathit{y}^*)\right] .
\end{equation}
%
%
%
%
%
Here $\Omega = \delta\omega/\Gamma$, with $\delta \omega = \omega_F - \omega_0$ being the frequency detuning of the drive from the mode eigenfrequency; 
it is assumed to be small for resonant driving, $\omega_0, \omega_F \gg |\delta \omega|$. 
Parameter $\beta_{\mathit{y}}$ is the scaled driving strength and $\xi_{\mathit{y}}(t)$ is Gaussian noise,
%
%
%
%
%
\begin{equation}
\label{eq:parameter}
\beta_{\mathit{y}}  = \frac{3 \gamma F^2}{32\omega_F^3\Gamma^3} \, , \quad
\xi_{\mathit{y}}  = \sqrt{ \frac{3 \gamma }{8\omega_F^3\Gamma^3} } \,  \xi (t) e^{-i\omega_F t}  \, , \quad
\langle \xi_{\mathit{y}} (t)\xi_{\mathit{y}} ^*(t')\rangle =  \frac{4 \alpha}{\Gamma} \delta(t-t') \, , \quad
\alpha = \frac{3 \gamma k_BT}{8\omega_F^3\Gamma}  \, .
\end{equation}
The evolution of $\mathit{y} (t)$ occurs slowly, on a timescale of $\sim \Gamma^{-1}$. On such timescale the correlator 
$\langle \xi_{\mathit{y}} (t)\xi_{\mathit{y}} (t')\rangle$, which is fast oscillating in time, can be disregarded, i.e., 
$\xi_{\mathit{y}} (t)$ is correlated only with $\xi_{\mathit{y}} ^*(t)$. 
The function $h_{\mathit{y}} $ is the scaled Hamiltonian $h$ introduced in Eq.~(4) of the main text, see also Fig.~1c.

In the absence of noise, the stationary  solutions of Eq.~(\ref{eq:Duffing-RWA-u})  
give the scaled complex amplitude of forced vibrations  $\mathit{y} _{j}$, where ${j}$ enumerates the solutions,
%
%
%
%
%
\begin{equation}
\label{eq:Amp} 
\mathit{y}_{{y}}  =-\frac{i\beta_{\mathit{y}}^{1/2}}{1+i \Omega - i {\left| \mathit{y}_{{j}} \right|}^2}  \, , 
\end{equation}
or
\begin{equation}
\phi_{\mathit{y}}(|\mathit{y}_{{j}}|^2) =0  \, , 
\quad 
\phi_{\mathit{y}}(x)= x\left[ 1+ {\left( x - \Omega \right)}^2 \right] - \beta_{\mathit{y}}
\end{equation}
The cubic equation $\phi_{\mathit{y}}(|\mathit{y}_{{j}}|^2) =0$ can have three real roots. 
This occurs in the region of vibration bistability, which is bounded by the bifurcational values of the dimensional parameter $\beta_{\mathit{y}}$
%
%
%
%
%
\begin{equation}
\label{eq:bet_bif}
(\beta_{\mathit{y}})_{B1,B2} = \frac{2}{27}[\Omega(\Omega^2 + 9) \pm (\Omega^2-3)^{3/2}] \, .
\end{equation}
The smallest and the largest  roots $|\mathit{y}_{{j}}|^2$ correspond to the stable vibrational states of the driven mode 
(denoted in the main text by lo and hi, respectively), the intermediate root corresponds to an unstable stationary state. 
The squared resonator amplitudes are  
$A_{{j}}^2 =(8\omega_F\Gamma/3\gamma)^{1/2} {\left| \mathit{y}_{j}\right|}^2$ with ${j}=\mathrm{hi},\mathrm{lo}$.  
An example of the dependence of $A_{{j}}$ on the drive frequency is shown as the response function in Fig.~1b of the main text.

%
%
%
\subsection{Linearized dynamics}
The noise leads to fluctuations of the driven resonator.
For weak noise the major effects are small-amplitude fluctuations about the stable vibrational states 
and rare events where the noise causes transitions between the states \cite{Dykman:1979}. 
A simple approach to the analysis of the small-amplitude fluctuations is based on linearizing the equation of motion Eq.~(\ref{eq:Duffing-RWA-u}) about the stable vibrations states \cite{Dykman:1979,Drummond:1980c,Dykman:1994gl}. 
The linearized equations for $\delta \mathit{y}_{{j}}(t) = \mathit{y}(t)-\mathit{y}_{j}$ read 
%
%
%
%
%
\begin{equation}
\label{eq:linear_dv}
\delta \dot{\mathit{y}}_{{j}}
= 
-\Gamma \left(
1 + i   \Omega -  2 i   {\left| \mathit{y}_{j} \right|}^2   
\right) \, \delta \mathit{y}_{{j}}
 + i  \Gamma \mathit{y}_{j}^2   \, \delta \mathit{y}_{{j}}^* 
- i \Gamma \xi_{\mathit{y}}  \, .
\end{equation}
The dynamics of $\delta \mathit{y}_{{j}}(t)$ 
is similar to the dynamics of a linear damped harmonic oscillator subject to noise. 
The characteristic frequency scale  of the fluctuations is determined by the eigenvalue of Eq.~(\ref{eq:linear_dv}) $\Gamma \nu_{ \mathit{y}, {j}}$,  
%
%
%
%
%
\begin{equation}
\label{eq:nu_0}
 \nu_{ \mathit{y}, {j}}^2 
 =   \left( 3 {\left| \mathit{y}_{j}  \right|}^2 - \Omega \right)\left( {\left| \mathit{y}_{j} \right|}^2 - \Omega \right)  + 1
= {\left( \omega_{{j}}/\Gamma \right)}^2  \, +  \,1  \,  
\end{equation}
%
%
%
%
with $\omega_{{j}}$ given by Eq.~(5) of the main text. 
We note that $\nu_{\mathit{y},{j}} \approx \omega_{{j}}/\Gamma$ 
for $\omega_{{j}} \gg \Gamma$.

%
%
%
%
\subsection{The power spectrum}
Fluctuations of the complex amplitude $y(t)$ lead to fluctuations of the coordinate $q(t)$ of the nanoresonator. 
Such fluctuations are of considerable broad interest, given that the driven nanomechanical resonator is a system far away from thermal equilibrium. 
They can be directly measured in the experiment by measuring the power spectrum of the nanoresonator. 
For the periodically driven nanoresonator  the power spectrum is defined as
%
%
%
%
%
\begin{align}
\label{eq:power_spectrum_defined}
&Q(\omega) =  \frac{1}{\pi}{\rm Re}\,\int_0^\infty \!\!\!\! dt \,\, e^{i\omega t}\bar Q(t), \nonumber\\
&\bar Q(t) = \frac{\omega_F}{2\pi}\int_0^{2\pi/\omega_F} \!\!\!\!\!\!\!\!\!\! dt' \,\,\left[\langle q(t+t')q(t')\rangle - \langle q(t+t')\rangle \langle q(t')\rangle\right].
\end{align}
Since the dynamics of the nanoresonator is mostly oscillations at frequency $\omega_F$ with the amplitude and phase that vary slowly over time $\sim 1/\omega_F$, the spectrum $Q(\omega)$ has peaks centered in a range $\sim |\delta\omega|, \Gamma$ around the drive frequency $\omega_F$ (and $-\omega_F$, strictly speaking).

Near $\omega_F$ the spectrum $Q(\omega)$ is given by the power spectrum of the complex amplitude ${\rm y}(t)$, as seen from Eq.~(\ref{eq:q_gamma}),
%
%
%
%
%
\begin{align}
\label{eq:spectrum_v_general}
Q(\omega) \approx 
\frac{2\omega_F\Gamma}{3\pi \gamma}\,{\rm Re}\,\int_0^\infty \!\!\!\!\!\! dt \,\, 
e^{i(\omega-\omega_F)t}\left\langle [\mathit{y}^*(t)-\langle \mathit{y}^*\rangle]\,[\mathit{y}(0)- \langle \mathit{y}\rangle]\right\rangle
\, . 
\end{align}
In the range of bistability, $\langle \mathit{y} \rangle$ is approximately the sum of the values 
$\mathit{y}_{\mathrm{ hi, lo}}$ weighted with the mean occupation  of the corresponding states. 
For weak noise, these occupations are of the same order of magnitude only in a very narrow parameter range (the region of the kinetic phase transition \cite{Dykman:1979}). 
Away from this region only one of the stable states is mostly occupied. 
Also, if the measurement is done over a time which is small compared to the reciprocal rate of interstate transitions, $\langle \mathit{y} \rangle$ should be 
replaced by the value $\mathit{y}_{{j}}$ in the state in which the system was prepared. 
Then $Q(\omega)$ is given by the partial spectrum 
%
%
%
%
%
\begin{align}
\label{eq:partial_spectrum_defined}
Q_{{j}}(\omega) = \frac{2\omega_F\Gamma}{3\pi \gamma}\,{\rm Re}\,
\int_0^\infty  \!\!\!\!\!\! dt \,\,  
e^{i(\omega-\omega_F)t}\left\langle \delta\mathit{y}_{{j}}^*(t)  \,
\delta\mathit{y}_{{j}}(0) 
\right\rangle \, . 
\end{align}
For a long observation time, in addition to the partial spectra given by Eq.~(\ref{eq:partial_spectrum_defined}) the full spectrum $Q(\omega)$ displays an extremely narrow peak in the region of the kinetic phase transition \cite{Dykman:1994gl}, which has been seen in micromechanical resonators driven by a noise that imitated a relatively high-temperature thermal noise \cite{Stambaugh:2006kg}. 
In this work we do not consider this peak, since our primary interest focuses on the spectra $Q_{{j}}(\omega)$.

It is straightforward to calculate $Q_{{j}}(\omega)$ in the region where the motion around the stable states can be linearized and is described by Eq.~(\ref{eq:linear_dv}). Such calculation was done in the quantum~\cite{Drummond:1980c} and classical \cite{Dykman:1994gl} theory, giving
%
%
%
%
%
\begin{equation}
\label{eq:spectrum}
Q_{{j}} (\omega) =  \frac{\Gamma k_BT}{2\pi\omega_F^2}\;
\frac{
{\left( \omega-\omega_F +2 \Gamma {\left| \mathit{y}_{{j}} \right|}^2  - \delta\omega  \right)}^2 + \Gamma^2 \left( 1 + {\left| \mathit{y}_{{j}}  \right|}^4 \right)
}
{
{\left[  {\left(\omega-\omega_F\right)}^2 - \Gamma^2 \nu_{ \mathit{y},{j}}^2 \right]}^2 
+ 4 \Gamma^2 {\left(\omega-\omega_F\right)}^2   
}
\, .
\end{equation}
The total area of the power spectrum Eq.~(\ref{eq:spectrum}) is 
%
%
%
%
%
\begin{align}
\label{eq:area_total}
{\mathcal I}_{{j}}
\equiv \int \!\! d\omega \,\, Q_{{j}}(\omega) = 
\frac{k_BT}{2\omega_F^2} \left[1 + \frac{| \mathit{y}_{{j}}|^4}{\nu_{\mathit{y},{j}}^2}\right] \, . 
\end{align}
It exceeds the area of the power spectrum in the absence of the drive, which is given by the above expression with $|\mathit{y}_{{j}}|^2$ set equal to zero. 
This facilitates the observation of
the satellite peaks of the power spectrum of the driven resonator even 
where in the absence of the driving the resolution of the experiment did not allow us to reliably detect the power spectrum.

For the discussion of the squeezing effect it is convenient to consider the areas of the power spectrum for
$\omega >\omega_F$ and for $\omega < \omega_F$, which we denote as ${\mathcal I}_{{j}}^{(+)}$ and ${\mathcal I}_{{j}}^{(-)}$, respectively. 
They have the form
%
%
%
%
%
%
\begin{align}
\label{eq:areas-results}
{\mathcal I}^{(\pm)}_{{j}} 
&= 
\frac{1}{2}{\mathcal I}_{{j}} \pm \Delta {\mathcal I}_{{j}} \, ,  
\qquad
\Delta {\mathcal I}_{{j}}
=
\frac{k_BT}{2\pi \omega_F^2}\,\,\frac{2| \mathit{y}_{{j}}|^2-\Omega}{\sqrt{\nu_{\mathit{y}, {{j}}}^2-1}}\,\arctan\left( \sqrt{\nu_{\mathit{y},{j}}^2- 1 } \right).
\end{align}
This expression can be analytically continued in a standard way to the region $\nu_{\mathit{y},{j}} <1$. 
The definition of the  areas in Eq.~(\ref{eq:areas-results}) takes into account the overlapping, for not too large $\nu_{{j}}$, of the satellite peaks
centered at $\approx \pm \Gamma\nu_{{j}}$. 
This analytic expression is used in the comparison with the experiment in Figs. 2 and 3 of the main text.

\begin{figure}[t!]
	\begin{center}
		\includegraphics[width=0.8\columnwidth]{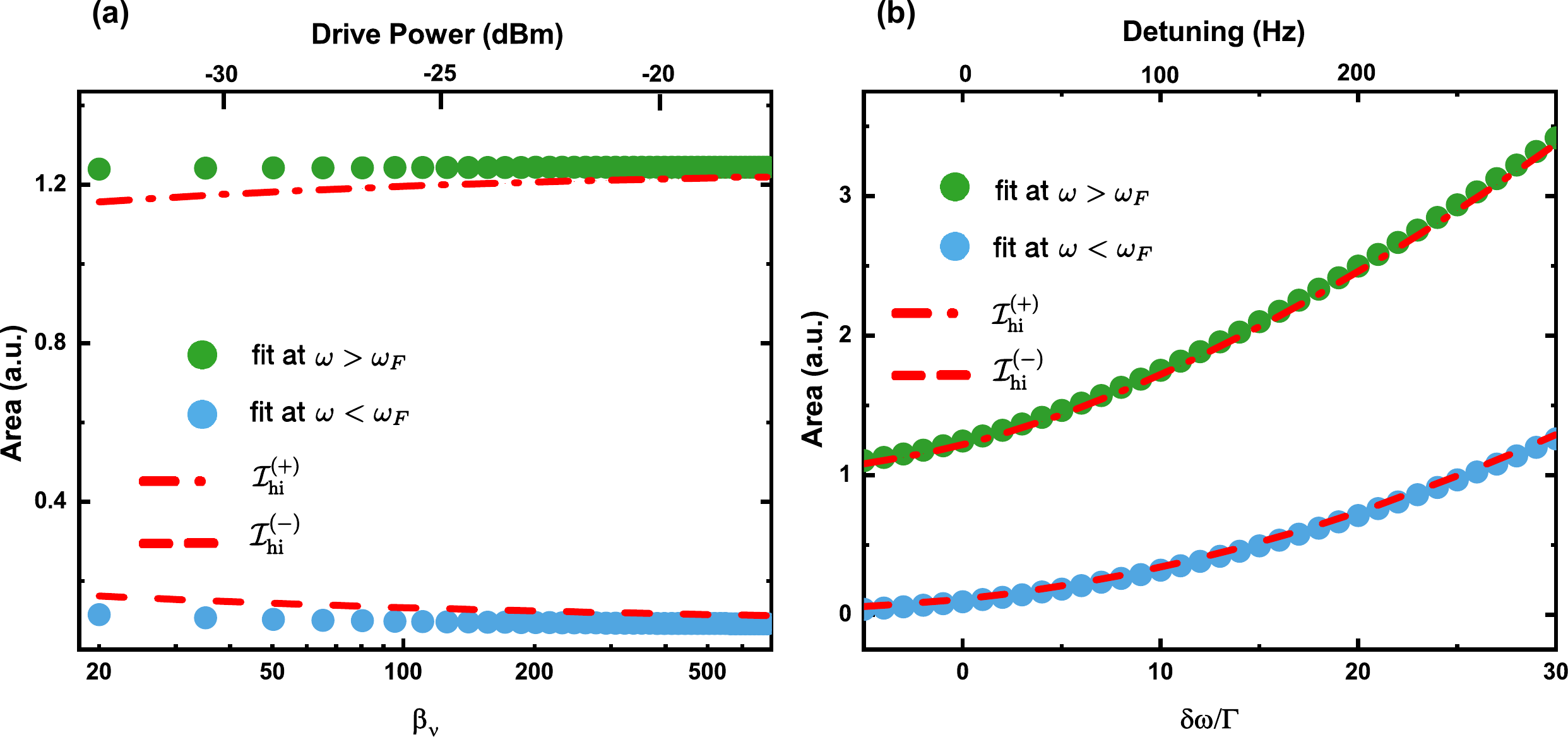} 
	\end{center}
	\caption{ 
		The red dash-dotted and dashed lines show the integrals of Eq.~(\ref{eq:areas-results}) 
		for the high amplitude state,  ${\mathcal I}^{(+)}_{\mathrm{hi}} $ and ${\mathcal I}^{(-)}_{\mathrm{hi}} $, respectively.
		The integrals are scaled  by $k_BT /(2 \omega_F^2)$. 
		The green and blue dots are obtained by fitting the power spectrum  $Q_{\mathrm{hi}} (\omega)$  of Eq.~(\ref{eq:spectrum}) (scaled by  $k_BT /(2 \omega_F^2)$) 
		by two Lorentzians centered at the maxima of $Q_{\mathrm{hi}} (\omega)$.
		{\bf a} Comparison as a function of scaled driving strength $\mathcal{\beta}_{\mathit{y}}$ at zero detuning $\delta\omega_F = 0$. 
		{\bf b }  Comparison as a function of scaled detuning $\delta\omega_F/\Gamma$ at $\mathcal{\beta}_{\mathit{y}}=700$.
	}
	\label{fig:S0-theory}
\end{figure}

The results for the two areas of Eq.~(\ref{eq:areas-results}) are shown in Fig.~\ref{fig:S0-theory} for the high amplitude state.
Here we compare these analytic results with 
the results of the following procedure, which is also the procedure used to extract the satellite areas from the experimental data. 
The power spectrum  $Q_{{j}} (\omega)$, Eq.~(\ref{eq:spectrum}),   is fit by two Lorentzians, with the maxima in the regions  $\omega>\omega_F$ and   $\omega<\omega_F$, respectively. 
The areas of the satellite peaks is then found from the parameters of the Lorentzians.
Figure~\ref{fig:S0-theory}a and b display these  areas as a function of the scaled driving strength $\mathcal{\beta}_{\mathit{y}}$ 
and the scaled detuning $\delta\omega_F/\Gamma$, respectively. 
The upper $x$-axis converts these values into the real drive power and detuning, for the parameters of the nanomechanical resonator studied in the experiment. 
The parameters are chosen to match those discussed in Figs.~2b and c, as well as Fig.~3c of the main text. 
There is a good agreement between the 
areas of the satellite peaks found from the Lorentzian fits and the areas of the parts of the analytical spectrum at $\omega>\omega_F$ and $\omega<\omega_F$ given by Eq.~(\ref{eq:areas-results}). 
%

%
%
%
%
%
%
%

%
%
%
\subsection{Squeezed thermal fluctuations for weak damping}
The power spectrum $Q_{{j}}(\omega)$ has a particularly simple form in the case of weak damping, where $\nu_{\mathit{y},{j}} \gg 1$. 
Here $Q_{{j}}(\omega)$ has two distinct, non-overlapping satellite peaks at frequencies $\omega_F \pm \Gamma \nu_{\mathit{y},{j}}$. 
Similar peaks emerge in the response of the resonator to an extra drive \cite{Dykman:1979}. 
%
%
%
It should be emphasized that the weak-damping condition implied here is much stronger than the condition $\Gamma\ll \omega_0$ of the nanoresonator to be underdamped.
The advantageous feature of the resonator studied in this work is that both conditions are met in the experiment, in a certain range of the amplitude and frequency of the driving field.

To analyze the dynamics in the weak-damping limit we re-write Eqs.~(\ref{eq:linear_dv}) separating the terms of the first order in the friction coefficient $\Gamma$. 
From Eq.~(\ref{eq:Amp}), to the leading order in $\Gamma$ we have  $\mathit{y}_{{j}}^2\approx \bar{\mathit{y}}_{{j}}^2$ 
with a real positive  $\bar{\mathit{y}}_{{j}}^2 \propto \Gamma^{-1}$ given by Eq.~(\ref{eq:bar_v}) below. 
To the first order in $\Gamma$
%
%
%
%
%
%
%
\begin{align}
\label{eq:bar_v}
\mathit{y}_{{j}} \approx \bar{\mathit{y}}_{{j}} 
\left(1 -\frac{i}{\bar{\mathit{y}}_{{j}}^2 -\Omega}\right), 
\qquad  
\bar{\mathit{y}}_{{j}}^2  {\left( \bar{\mathit{y}}_{{j}}^2 - \Omega \right)}^2 =\beta_{\mathit{y}} \, .
\end{align}
Equations~(\ref{eq:linear_dv}) for the small 
deviations $\delta \mathit{y}_{{j}}, \delta \mathit{y}_{{j}}^*$ from the stable states can be written as
%
%
%
%
%
%
%
\begin{equation}
\label{eq:dv_Hamiltonian}
\delta\dot{\mathit{y}}_{{j}} 
= i\,\frac{\partial \delta h_{{j}}}{\partial \delta \mathit{y}_{{j}}^*} - \Gamma(\delta \mathit{y}_{{j}}+\mu_{{j}} \delta \mathit{y}_{{j}}^*) -i\Gamma\xi_{\mathit{y}}(t) 
\, , \qquad 
\delta h_{{j}} 
= \Gamma( 2\bar{\mathit{y}}_{{j}}^2-\Omega)|\delta \mathit{y}_{{j}}|^2 
+\frac{1}{2}\bar{\mathit{y}}_{{j}}^2 \left [ (\delta \mathit{y}_{{j}}^*)^2 + \delta \mathit{y}_{{j}}^2\right] \, . 
 \end{equation}
Here, the Hamiltonian $\delta h_{{j}}$ 
is the expansion of the Hamiltonian $h_{\mathit{y}}$, 
Eq.~(\ref{eq:Duffing-RWA-u}), about  $\bar{\mathit{y}}_{{j}}$ to the second order 
in $\delta \mathit{y}_{{j}}, \delta \mathit{y}_{{j}}^*$. 
Parameter $\mu_{{j}} =2\bar{\mathit{y}}_{{j}}^2/(\bar{\mathit{y}}_{{j}}^2-\Omega)$ was introduced in the main text. 
It should be noted that the phase of the scaled complex amplitude  $\mathit{y}_{{j}}$ in a stable state ${{j}}$ is ``pinned'' to the phase of the driving field. 
This phase directly reflects the broken time-translation symmetry of the stable state. 
The Hamiltonian $\delta h_{{j}}$ thus also ``knows'' about the phase of the field. 
This leads to the terms $\propto (\delta \mathit{y}_{{j}})^2, (\delta \mathit{y}_{{j}}^*)^2$. 
If there were no such terms, the quadratures of the fluctuations about the stable state would be 
equal (we recall that these quadratures are given by the scaled Re~$\delta \mathit{y}_{{j}}$ and $-{\rm Im}~\delta \mathit{y}_{{j}}$, respectively). 
Indeed, as seen from the expression for the correlators $\xi_{\mathit{y}}(t)$, in this case we would have 
$\langle \delta \mathit{y}_{{j}}^2\rangle = \langle (\delta \mathit{y}_{{j}}^*)^2\rangle =0$. 
However, the latter expression does not apply to our driven nonlinear resonator.

Since, as noted above, for weak damping $\mathit{y}_{{j}} \propto \Gamma^{-1/2}$  and $\Omega \propto \Gamma^{-1}$, 
the Hamiltonian $\delta h_{{j}}$ is independent of $\Gamma$. 
This Hamiltonian can be diagonalized by the Bogoliubov (squeezing) transformation
%
%
%
%
%
%
%
\begin{align}
\label{eq:squeezing_transform}
\delta \mathit{y}_{{j}} = u_{{j}} \cosh \varphi_{{j}} - u_{{j}}^* \sinh \varphi_{{j}} \, ,
\end{align}
so that the new Hamiltonian for the variables $u_{{j}}$ contains only $|u_{{j}}|^2$ and not $u_{{j}}^2$, $(u_{{j}}^*)^2$. 
This condition is met if we set
%
%
%
%
%
%
%
\begin{align}
\label{eq: squeezing_phase}
\tanh \left( 2\varphi_{{j}} \right) 
 = \frac{\bar{\mathit{y}}_{{j}}^2}{2\bar{\mathit{y}}_{{j}}^2-\Omega}  \, .
\end{align}
The resulting equation of motion is
%
%
%
%
%
%
%
\begin{equation}
\label{eq:squeezed_Hamiltonian}
\dot u_{{j}} = 
i\,\frac{\partial \delta h_{{j}}}{\partial u_{{j}}^*} - \Gamma u_{{j}} - \Gamma \mu_{{j}} u_{{j}}^* 
-i\Gamma\xi_{{j}}(t)
\, ,
\quad
\delta h_{{j}} 
= 
{\cal S}_{{j}} 
\Gamma \nu_{\mathit{y},{j}} |u_{{j}}|^2\,, 
\quad {\cal S}_{{j}}= \mathrm{sgn}(2\bar{\mathit{y}}_{{j}}^2-\Omega) \, , 
\qquad \xi_{{j}}(t) = \xi_{\mathit{y}}(t) \cosh \varphi_{{j}} - \xi^{*}_{\mathit{y}}(t)\sinh \varphi_{{j}} \, .
\end{equation}
In this expression only the leading-order term with respect to $\Gamma$  should be taken into account in  
$\nu_{\mathit{y},{j}}$, i.e., $\nu_{\mathit{y},{j}} = [(3\bar{\mathit{y}}_{{j}}^2-\Omega)(\bar{\mathit{y}}_{{j}}^2 -\Omega)]^{1/2}$. 
Clearly, $\Gamma \nu_{\mathit{y},{j}}$ is independent of $\Gamma$. 
The coefficient ${\cal S}_{{j}} =\pm 1$ is given by the sign of  $2\bar{\mathit{y}}_{{j}}^2 -\Omega$. 
Parameter $\varphi_{{j}}$ is simply related to the standard~\cite{Walls2008} parameters $(r,\theta)$  
of the squeezing transformation: $r=\varphi_{{j}}, \theta =0$ for ${\cal S}_{{j}}=1$, and 
$r=|\varphi_{{j}}|, \theta =\pi$ for ${\cal S}_{{j}}=-1$.

To the lowest order in $\Gamma$, the variable $u_{{j}}$ is oscillating as $\exp[i{\cal S}_{{j}} \Gamma \nu_{\mathit{y},{j}} t]$.
The coefficient ${\cal S}_{{j}}$ reflects the fact that, in the range of bistability, $u_{{j}}$ is rotating in the opposite directions 
for the large and small amplitude stable states where, as it is well-known for the Duffing resonator, $2\bar{\mathit{y}}_{{j}}^2 -\Omega$ 
is positive or negative, respectively. 
Function $u_{{j}}^*(t)$ is rotating in the direction opposite to $u_{{j}}(t)$. 
Therefore, for small $\Gamma$ (or equivalently, for $\nu_{\mathit{y},{j}} \gg 1$), the term $\propto \mu_{{j}}$ 
in Eq.~(\ref{eq:squeezed_Hamiltonian}) can be disregarded. 
Then, using the Stratonovich convention for averaging $\delta$-correlated noise, we see from Eqs.~(\ref{eq:parameter}) and (\ref{eq:squeezed_Hamiltonian}) that
%
%
%
%
%
%
%
\begin{equation}
\langle |u_{{j}}|^2\rangle \approx 2\alpha\cosh 2\varphi_{{j}}\, , 
\end{equation}
whereas the mean value $\langle u_{{j}}^2\rangle \approx -i(2\alpha/\nu_{\mathit{y},{j}})\sinh 2\varphi_{{j}}$ is much smaller and can be disregarded. 
As a result we have for the average values of the quadratures
%
%
%
%
%
%
%
\begin{align}
\label{eq:complex_quadratures}
\langle ({\rm{Re}}\,\delta \mathit{y}_{{j}})^2\rangle \approx \frac{\alpha}{2}\left(1+e^{-4\varphi_{{j}}}\right), 
\qquad \langle ({\rm{Im}}\,\delta \mathit{y}_{{j}})^2\rangle \approx \frac{\alpha}{2}\left(1+e^{4\varphi_{{j}}}\right)
\, , 
\end{align}
which demonstrates the squeezing of thermal fluctuations about the stable state of forced vibrations. Going back to the unscaled
 quadratures $\mathit{x}_1,\mathit{x}_2$, Eq.~(\ref{eq:complex_quadratures}) gives
%
%
%
%
%
%
%
\begin{equation}
\label{eq:quadratures_dimensional}
\langle \delta \mathit{x}^2_1 \rangle =  \frac{k_BT}{2 \omega^2_F}  \, \left( 1 + e^{- 4\varphi_{{j}}} \right)
\, , \qquad 
\langle\delta \mathit{x}^2_2  \rangle =  \frac{k_BT}{2 \omega^2_F}  \, \left( 1 + e^{ 4\varphi_{{j}} } \right) \, .
\end{equation}
\noindent
Equation~(\ref{eq:squeezed_Hamiltonian}) for $u_{{j}}$  has a simple solution
%
%
%
%
%
%
%
\begin{align}
\label{eq:explicit_squeezed}
u_{{j}}(t) 
= 
\exp[-(\Gamma-i{\cal S}_{{j}} \Gamma \nu_{\mathit{y},{j}}) t]\,
\left[u_{{j}}(0)-i\Gamma\int_0^t \!\!\!\!\! dt'\,\exp[(\Gamma-i{\cal S}_{{j}} \Gamma \nu_{\mathit{y},{j}}) t'] \,\, 
 \xi_{{j}}(t')\right] \, .
\end{align}
With this solution, using Eqs.~(\ref{eq:spectrum_v_general}) and (\ref{eq:squeezing_transform}) we obtain
%
%
%
%
%
%
%
\begin{align}
Q_{{j}} (\omega) 
& \approx 
\frac{\Gamma k_BT}{4\pi\omega_F^2} 
\,\, 
\frac{\cosh 2\varphi_{{j}} \, (1+\cosh 2\varphi_{{j}})}{(\omega-\omega_F -{\cal S}_j\Gamma\nu_{\mathit{y},{j}})^2 + \Gamma^2}
\,  \quad \mbox{for} \quad
\left|\omega-\omega_F -{\cal S}_{{j}} \Gamma\nu_{\mathit{y},{j}} \right| \ll \Gamma\nu_{\mathit{y},{j}} \, ,  \nonumber \\
Q_{{j}} (\omega) 
& \approx 
\frac{\Gamma k_BT}{4\pi\omega_F^2}
\,\,
\frac{\cosh 2\varphi_{{j}} \, (\cosh 2\varphi_{{j}}-1)}{(\omega-\omega_F +{\cal S}_{{j}}\Gamma\nu_{\mathit{y},{j}})^2 
+ \Gamma^2}
\,  \quad \mbox{for} \quad
\left|\omega-\omega_F +{\cal S}_{{j}} \Gamma\nu_{\mathit{y},{j}} \right| \ll \Gamma\nu_{\mathit{y},{j}}
\, . 
\label{eq:spectrum_squeezed}
\end{align}
The power spectrum Eq.~(\ref{eq:spectrum_squeezed}) coincides with Eq.~(11) of the main text if one takes into account that 
${\cal S}_{{j}}=1$ for the high amplitude stable state and ${\cal S}_{{j}}=-1$ for the low amplitude stable state. 
The spectrum consists of two Lorentzian peaks with half width  $\Gamma$. 
The peaks are located at the drive frequency shifted up and down by the frequency of vibrations in the rotating frame about the stable state ${{j}}$. 
We emphasize that, since we kept only the leading-order terms in the decay rate $\Gamma$,  
Eq.~(\ref{eq:spectrum_squeezed}) applies only in the frequency ranges of the peaks.

It is instructive also to compare the areas of the satellite peaks described by Eq.~(\ref{eq:spectrum_squeezed}) 
with Eq.~(\ref{eq:areas-results}) for the areas of the peaks on the opposite sides of $\omega_F$. 
From  Eq.~(\ref{eq:spectrum_squeezed}), the areas of the peaks are
%
%
%
%
%
%
%
\begin{equation}
{\cal I}_{{j}}^{(\pm)} = 
\frac{k_BT}{4\omega_F^2}
\cosh 2\varphi_{{j}}
 \, (\cosh 2\varphi_{{j}}  \pm 1)
 =
 \frac{k_BT}{4\omega_F^2}
 \left[1+\frac{\bar{\mathit{y}}_{{j}}^4}{\nu_{\mathit{y},{j}}} \pm \frac{2\bar{\mathit{y}}_{{j}}^2-\Omega}{ \nu_{\mathit{y},{j}} } \, {\cal S}_{{j}}\right] \, .
\end{equation}
One can easily see that this expression coincides with Eq.~(\ref{eq:areas-results}) if in the latter equation one goes 
to the limit $\nu_{\mathit{y},{j}} \gg 1$ and replaces $|\mathit{y}_{{j}}|^2$ with $\bar{\mathit{y}}_{{j}}^2$, which corresponds to keeping 
the terms of the lowest-order in $\Gamma$.

The central feature of the spectrum  Eq.~(\ref{eq:spectrum_squeezed}) is that the two satellite peaks of $Q_{{j}}(\omega)$ have different amplitudes and areas. 
This is a direct indication of the squeezing of the vibrations about the stable state. We emphasize that this is not the cause of the squeezing. Both the difference in the areas and the squeezing are a consequence of the broken time translation symmetry in a periodically driven system, and because of that they are immediately related to each other. This is why measuring the ratio of the peak heights allows one to directly determine the squeezing parameter $\varphi_{{j}}$ in the underdamped system and thus, 
if we use the conventional notations of quantum optics, the squeezing parameter $r=|\varphi_{{j}}|$.

%
\subsection{Asymmetry of the satellite peaks }
In calculating the spectrum  $Q_{{j}}(\omega)$ in the previous section we expanded the Hamiltonian $h_{\mathit{y}}$ to second order in $\delta \mathit{y}, \delta \mathit{y}^*$. 
Disregarding higher-order terms may be inconsistent in the limit of  small decay rate $\Gamma$. 
These terms describe the nonlinearity of the vibrations about the stable states in the rotating frame. 
Such nonlinearity leads to the dependence of the vibration frequency on the vibration amplitude. 
Thermal fluctuations of the amplitude then translate into thermal fluctuations of the vibration frequency 
and thus lead to broadening of the satellite peaks. 
Such broadening is well understood for nonlinear resonators in the absence of strong periodic driving \cite{Dykman:1984}.

A quantum theory of the nonlinearity-induced spectral broadening of the peaks of $Q_{{j}}(\omega)$ was discussed in 
Ref.~[\citenum{Dykman2012}]. 
Here we will briefly outline the corresponding classical theory. 
As indicated in the main text, the amplitude dependence of the vibration frequency comes from the Duffing nonlinearity in the first order of the perturbation theory. 
More generally, it is well-known from classical mechanics~\cite{Arnold:1989} that the first-order terms in the amplitude dependence of the vibration frequency come from the terms in the Hamiltonian that are nonzero when averaged over the vibration period. 
In the case of the Duffing resonator the corresponding term is $\gamma q^4/4$ which, for $q=A\cos(\omega_0 t+\phi)$, has the period-averaged value $3\gamma A^4/32$. 

In line with the above argument, to find the amplitude dependence of the vibrations in the rotating frame about a ${j}$th state one has to do the following steps: 
\begin{itemize}
	\item Expand the full Hamitlonian $h_{\mathit{y}}$ about $\bar{\mathit{y}}_{{j}}$ to the 4th order in $\delta \mathit{y}_{{j}}, \delta \mathit{y}_{{j}}^*$.
	\item Express $\delta \mathit{y}, \delta \mathit{y}^*$ in terms of $u_{{j}},u_{{j}}^*$
	\item Out of all quartic terms in $u_{{j}}, u_{{j}}^*$ keep only the term $\propto {|u_{{j}}|}^4$, as this is the only term that does not oscillate in the harmonic approximation. 
\end{itemize}
The above routine has to be augmented to allow for the fact that the cubic terms 
$\delta h_{{j}}^{(3)} = \bar{\mathit{y}}_{{j}} {|\delta \mathit{y}_{{j}}|}^2\cdot ( \delta \mathit{y}_{{j}} +\delta \mathit{y}_{{j}}^*)$ 
will contribute to the amplitude dependence of the frequency when taken to the second order of the perturbation theory. 
The corresponding terms renormalize the coefficient in front of the term $\propto {|u_{{j}}|}^4$. 
The result is as if the relevant quartic term in $u_j,u_j^*$ of the  effective Hamiltonian has the form
%
%
%
%
%
%
%
\begin{equation}
\label{eq:h_eff}
\delta h_{{j}}^{(4)}= \frac{1}{2} \, \Gamma \, B_{{j}} \,  {\left| u_{{j}} \right|}^4
\, , \qquad 
B_{{j}} 
=
\frac{1}{2}
\left( 3\cosh^2 2\varphi_{{j}} -1 \right)
 -\frac{\bar{\mathit{y}}_{{j}}^2 \exp(-2\varphi_{{j}})}{{\cal S}_{{j}}\nu_{{j}}}
 \left(
 \frac{9}{2}\cosh 4\varphi_{{j}} - 3\sinh 4\varphi_{{j}} +\frac{3}{2}
 \right) \, .
\end{equation}
The parameter $B_{{j}}$ is determined by the ratio $\beta = \beta_{\mathit{y}}/\Omega^3$. 
A plot of $V=2 B_{{j}}$ and $\nu_0 = \nu_{{j}}/\Omega$ 
as functions of this ratio for the large- and small-amplitude attractor is shown in Fig.~9 of Ref.~[\citenum{Dykman2012}].  
The Hamiltonian $\delta h_{{j}}^{(4)}$ has to be added to $\delta h_{{j}}$ 
in the equation of motion for $u_{{j}}$, Eq.~(\ref{eq:squeezed_Hamiltonian}).
The resulting equation has the same general form as the equation of motion of an anharmonic Duffing resonator with eigenfrequency 
$\Gamma{\cal S}_{{j}} \nu_{{j}}$ 
in the presence of relaxation and noise, but with no driving. 
Using the results \cite{Dykman1971,Dykman2012} 
we can write the power spectrum of the resonator near frequency  $\omega_F+\Gamma{\cal S}_{{j}} \nu_{\mathit{y},{j}}$ 
in the form of a series
%
%
%
%
%
%
%
\begin{align}
\label{eq:spectrum_nonlinear}
Q_{{j}}(\omega)
&\approx \frac{k_BT}{4\pi\omega_F^2} \,
\cosh 2\varphi_{{j}} \left( \cosh 2\varphi_{{j}} +1 \right)  
\, {\rm Re} \, \sum_n \phi_{{j}} \left(n,\omega-\omega_F -\Gamma {\cal S}_{{j}} \nu_{\mathit{y},{j}} \right)
\, , \\
\phi_{{j}}(n,\omega')
&=
\frac{
4n(\Lambda_{{j}}-1)^{n-1}(\Lambda_{{j}}+1)^{-(n+1)} 
}
{ \Gamma(2\aleph_{{j}} n-1)-i\omega' } \, .
\end{align}
Here,
%
%
%
%
%
%
%
%
%
\begin{align}
\label{eq:spectral_notations}
\Lambda_{{j}} 
= (1+2i B_{{j}} \alpha\cosh2\varphi_{{j}})/\aleph_{{j}} \,  ,
\qquad 
\aleph_{{j}} = (1+4i B_{{j}}\alpha\cosh2\varphi_{{j}})^{1/2} 
\qquad [{\rm Re}\,\aleph_{{j}}>0]
\,.
\end{align}
Near frequency $\omega_F-\Gamma{\cal S}_{{j}}  \nu_{{j}} $ the power spectrum has the form
%
%
%
%
%
%
%
%
%
\begin{equation}
\label{eq:mirror_frequency}
Q_{{j}} (\omega)
\approx 
\frac{k_BT}{4\pi\omega_F^2}
\,\cosh 2\varphi_{{j}} \left( \cosh 2\varphi_{{j}}  -1 \right)
\, 
{\rm Re} \,\sum_n \phi_j[n,-(\omega-\omega_F +\Gamma{\cal S}_{{j}}  \nu_{\mathit{y},{j}} )] \, .
\end{equation}
The shape of the spectra in Eqs.~(\ref{eq:spectrum_nonlinear}) 
and (\ref{eq:mirror_frequency}) is determined by the parameter $B_{{j}}  \alpha$, i.e., by the effective nonlinearity of the vibrations about the stable vibrational state. 
If $|B_{{j}} |\alpha \ll1$, 
the main contribution to the spectra comes from the term $n=1$ 
and the spectra are close to the Lorentzian spectra of Eq.~(\ref{eq:spectrum_squeezed}). 
However, if $|B_{{j}} |\alpha \gtrsim 1$, 
the shape of the satellite peaks  strongly differs from the Lorentzian shape. %
Since $|B_{{j}} |\sim 1$, the important parameter of the spectral shape is $\alpha$. This parameter characterizes the ratio of the broadening of the spectrum due to thermal fluctuations of the vibration amplitude and the decay rate. Equations~(\ref{eq:spectrum_nonlinear}) and (\ref{eq:mirror_frequency}) allow one to calculate the shape of the spectrum for an arbitrary $\alpha$ and to see the evolution of the spectrum with increasing $\alpha$.

We wish to make the following two comments: First, the nonlinearity of the vibrations about the stable states, although it can dramatically change the shape of the peaks at frequencies $\omega_F\pm{\cal S}_{{j}} \Gamma\nu_{\mathit{y},{j}}$ for weak damping, does not change the area of these peaks, to the leading order. 
Second, it is important to keep in mind that Eq.~(\ref{eq:spectrum_squeezed}) applies in the weak damping approximation discussed in section I.D. 
In this approximation, the decay rate has to be small, so that the  satellite peaks  described by Eq.~(\ref{eq:spectrum_squeezed}) are well resolved. 
However, it should be not too small if one approximates the shape of the peaks by a Lorentzian.

\subsection{Characterizing sideband squeezing in the quantum regime}
The previous analysis referred to the classical regime, where squeezing is explicitly seen in the intensities of the sideband spectral peaks in the power spectrum. In the quantum regime, of primary interest is the  emission (fluorescence)  spectrum of the driven mode. It is determined by the correlation function of the ladder operators $a$ and $a^\dagger$. For the considered unit-mass mode, its shape away from $\omega_F$  is described by Eq.~(\ref{eq:power_spectrum_defined}) in which one should replace $q(t+t')$ with $(\hbar/2\omega_0)^{1/2}a^\dagger(t+t')$ and $q(t')$ with $(\hbar/2\omega_0)^{1/2}a(t')$. 

As indicated in the main text, it was noticed in Ref.~\onlinecite{Drummond:1980c} that in the ultra-quantum regime, where the Planck number $\bar{n} =[\exp(\hbar\omega_0/k_BT)-1]^{-1} $ is very small, $\bar n \ll 1$,
the sideband peaks in the emission spectrum of a resonantly driven mode have equal areas. This means that the squeezing is not seen in the emission spectrum, in contrast to the classical regime.

However, the squeezing itself exists. In the considered case of a strongly underdamped mode, using the results \cite{Dykman2012} one can show that the (unscaled) variances $\sigma^2_{{{j}}, {\rm in}}$ and $\sigma^2_{{{j}} {\rm ,quad}}$ of the in-phase and quadrature components of the mode displacement are
\begin{align}
\label{eq:variances}
 \sigma^2_{ {{j}}, \rm in} = \frac{\hbar}{4\omega_F}(2\bar n+1)(1+e^{-4\varphi_{{j}}}),
 \qquad \sigma^2_{ {{j}}, \rm quad} = \frac{\hbar}{4\omega_F}(2\bar n+1)(1+e^{4\varphi_{{j}}}).
 \end{align}
This expression can also be obtained from Ref.~\onlinecite{Buks2006a} in the limit of weak damping. 
In the classical limit $k_BT\gg \hbar\omega_0$ the expressions for  $\sigma^2_{ {j}, \rm in}$ and $\sigma^2_{ {j},\rm quad}$ go over into the expressions (\ref{eq:quadratures_dimensional}) for $\langle \delta x_1^2\rangle$ and $\langle\delta x_2^2\rangle$, respectively. In contrast, in the low-temperature limit $k_BT\ll \hbar\omega_0$ one sees that $\sigma^2_{ {j}, \rm in}= (\hbar/4\omega_F)[1+\exp(-4\varphi_{{j}})]$ is below the standard  quantum limit  $\hbar/2\omega_F$ for an undriven mode.

In the side-band resolved regime the squeezing can be revealed via a spectral measurement at any temperature by measuring the spectrum of the response of the mode to a weak probe driving field $F'\exp(-i\omega' t)$ with  the frequency $\omega'$ close to the frequency of the strong field  $\omega_F$. As indicated in the main text, such response can be characterized by the susceptibilities $\chi(\omega')$ and ${\cal X}(\omega')$ that relate the probe-induced  forced vibrations $\delta \langle q(t)\rangle$ to the  amplitude $F'$ of the probe field, $\delta\langle q(t)\rangle = \chi(\omega')F'\exp(-i\omega' t)+{\cal X}(\omega')F'\exp[-i(2\omega_F-\omega')t]$. These susceptibilities were discussed earlier \cite{Dykman:1979,Dykman:1994gl}). In the present notations, the susceptibility $\chi(\omega')$ of the mode in a state $j$ is
  \begin{align}
\label{eq:susceptibility}
\chi_{{j}}(\omega') = \frac{i}{2\omega_F}\frac{\Gamma -i(\omega'-\omega_F) + i(\delta\omega - 2\Gamma |{\rm y}_{{j}}|^2)}
{\Gamma^2\nu_{ \mathit{y},{j} }^2 - 2i\Gamma (\omega'-\omega_F) - (\omega'-\omega_F)^2}.
\end{align}
Of particular physical interest is the value of Im~$\chi_{ j}(\omega)$. On the one hand, it gives the quadrature of the displacement $\delta\langle q(t)\rangle$ at frequency $\omega'$. On the other hand, it determines how much energy the resonator absorbs from the probe field at frequency $\omega'$\cite{Landau1980}. It follows from Eq.~(\ref{eq:susceptibility}) that the integral of the absorption coefficient over the frequency is positive, $\int d\omega' \,{\rm Im}\,\chi_{ j}(\omega') =\pi/2\omega_F >0$, which is a well-known general property of a weakly nonlinear oscillator in a stationary regime. However, along with absorbing, a nonequilibrium mode can also amplify the probe field in the resonator (for example, amplify the electromagnetic field in a cavity) at the expense of the energy of the strong field $\propto F$. This happens in the frequency range where  Im~$\chi_{ j}(\omega')$ is negative.

Along with Im~$\chi_{{j}}(\omega')$, of interest is also the function $|\chi_{{j}}(\omega')|^2$, which determines the squared amplitude of the forced vibrations at the probe field frequency. In the sideband resolved regime, both Im~$\chi_{{j}}(\omega')$ and $|\chi_{{j}}(\omega')|^2$ have Lorentzian peaks at $\omega'-\omega_F = \pm \Gamma\nu_{ y,{j}}$ with halfwidth $\Gamma$
\begin{align}
\label{eq:susc_peaks}
&{\rm Im}~\chi_{{j}}(\omega') \approx \frac{1}{4\omega_F \nu_{ \mathit{y},{j}}}\,\frac{\Gamma \nu_{ \mathit{y},{j}} 
\pm (\delta\omega - 2\Gamma |{\rm y}_{{j}}|^2)}
{[(\omega'-\omega_F) \pm \Gamma \nu_{ \mathit{y},{j}} ]^2 + \Gamma^2},\nonumber\\
|\chi_{{j}}(\omega')|^2 \approx \frac{1}{16\omega_F^2\Gamma^2 \nu_{ \mathit{y},{j}}^2}\,&
\frac{[\Gamma \nu_{ \mathit{y},{j}} \pm (\delta\omega - 2\Gamma |{\rm y}_{{j}}|^2)]^2}
{[(\omega'-\omega_F) \pm \Gamma \nu_{ \mathit{y},{j}} ]^2 + \Gamma^2}, \qquad \bigl|\Gamma^2 \nu_{ \mathit{y},{j}}^2 -|\omega'-\omega_F|^2|\lesssim \Gamma^2
\end{align}
As seen from Eq.~(\ref{eq:susceptibility}), the function Im~$\chi_{{j}}(\omega')$ 
has a positive peak at frequency $\omega' \approx  \omega_F +{\cal S}_{{j}} \Gamma\nu_{\mathit{y},{j}}$, which corresponds to absorption of the probe field, and it  has a negative peak at $\omega' \approx \omega_F -{\cal S}_{{j}} \Gamma\nu_{\mathit{y},{j}}$, which corresponds to amplification of the probe field  \cite{Dykman:1979}.

From Eqs.~(\ref{eq: squeezing_phase}) and (\ref{eq:susc_peaks}) one can readily find that the ratio of the areas of the peak 
of  Im~$\chi_{{j}}(\omega')$ at $\omega' \approx \omega_F +{\cal S}_{{j}} \Gamma\nu_{\mathit{y},{j}}$ to the area of the peak of $|{\rm Im}~\chi_{{j}}(\omega')|$ at $\omega'\approx \omega_F -{\cal S}_{{j}} \Gamma\nu_{\mathit{y},{j}}$ is equal to $\tanh^2\varphi_{{j}}$, to the leading order in $1/\nu_{\mathit{y},{j}}$. 
Similarly, the ratio of the areas of the corresponding peaks of $|\chi_{{j}}(\omega')|^2$ is $\tanh^4\varphi_{{j}}$. These ratios are thus determined by the squeezing parameter, both in the classical and in the quantum regimes. This shows that fluctuation squeezing in a resonantly driven mode has a pronounced spectral manifestation not only in the classical, but also in the quantum regime.

%
%
%
%
\section{Experimental setup, chararcterization and calibration}
%
%
%
%
%
%
\subsection{Sample and experimental measurement setup}
\begin{figure}[h!]
	\begin{center}
	\includegraphics[width=0.4\columnwidth]{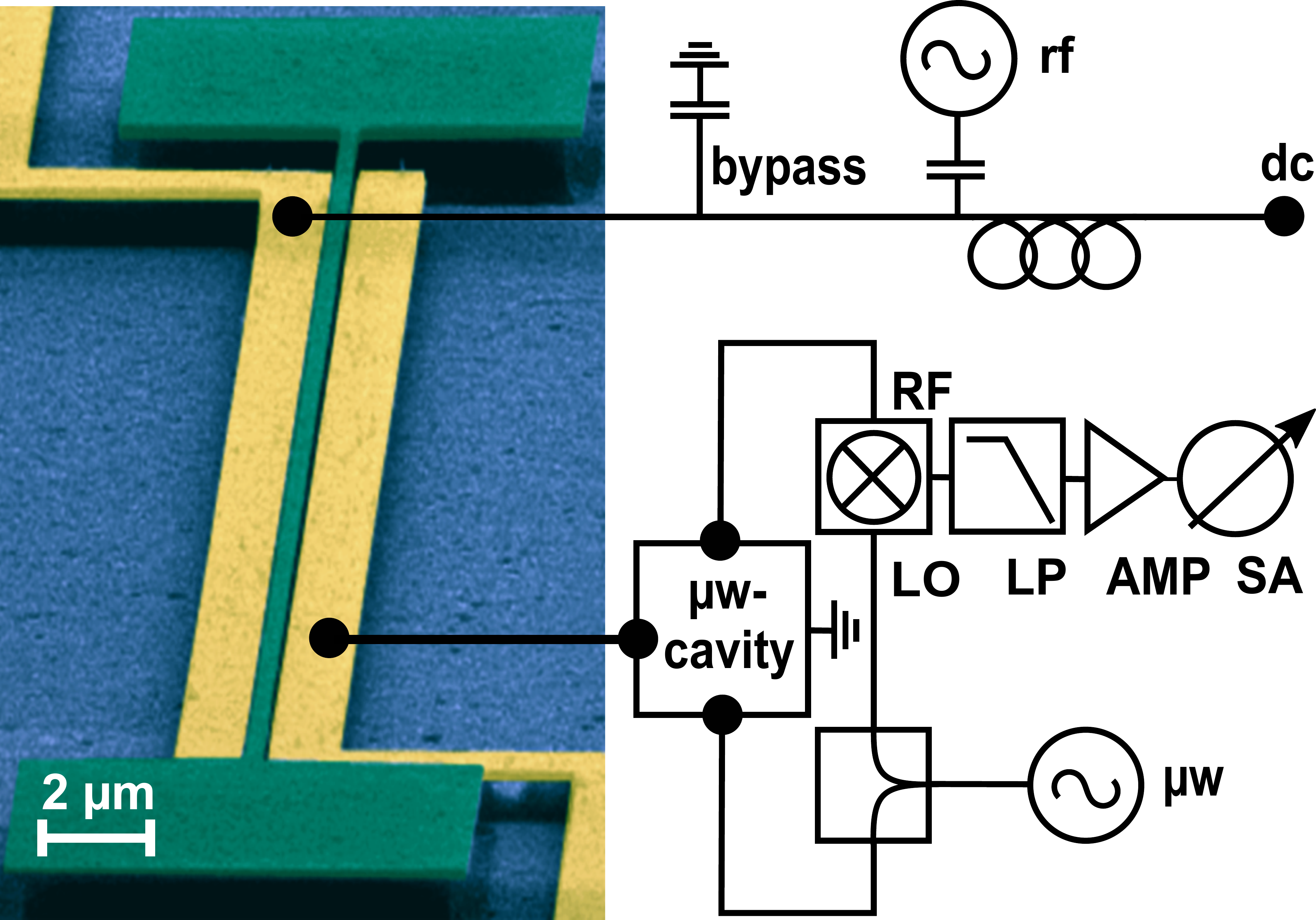}
	\end{center}
	\caption{ 
		 Simplified electronic set-up for dielectric actuation and displacement detection. 
			         }
	\label{fig:set-up}
\end{figure}

%
The nanomechanical resonator under investigation is a doubly clamped silicon nitride string resonator.
It is $270$\,nm wide, $100$\,nm thick and $55$\,\textmu m long and flanked by two adjacent gold electrodes for dielectric control. 
A schematic of the dielectric measurement setup is depicted in Fig.~\ref{fig:set-up}. 
A microwave signal (\textmu w) is used to resonantly pump the microwave cavity, which is bonded to one of the electrodes. 
The transmitted and modulated microwave cavity signal (RF) is demodulated by IQ-mixing it with a reference signal (LO). 
The output signal is low-pass filtered (LP), amplified (AMP) and its frequency spectrum is recorded using a spectrum analyzer (SA). 
Only for the measurement shown in Fig.~1b of the main text as well as Fig.~\ref{fig:characterization}a, c and Fig.~\ref{fig:banana} a fast lock-in amplifier is employed.
The dc voltage (dc) and the rf drive tone (rf) of frequency $f_F$ are combined with a bias tee and applied to the other electrode. 
A microwave bypass enables the combination of dielectric actuation and detection. 
More details about the dielectric control scheme can be found in Refs.~\citenum{Unterreithmeier:2009gh,Rieger2012,Faust:2012}.

%
%
%
%
%
%
%

%
%
%
\subsection{Calibration using the linear response function}
\begin{figure}[t]
	\includegraphics[width=\columnwidth]{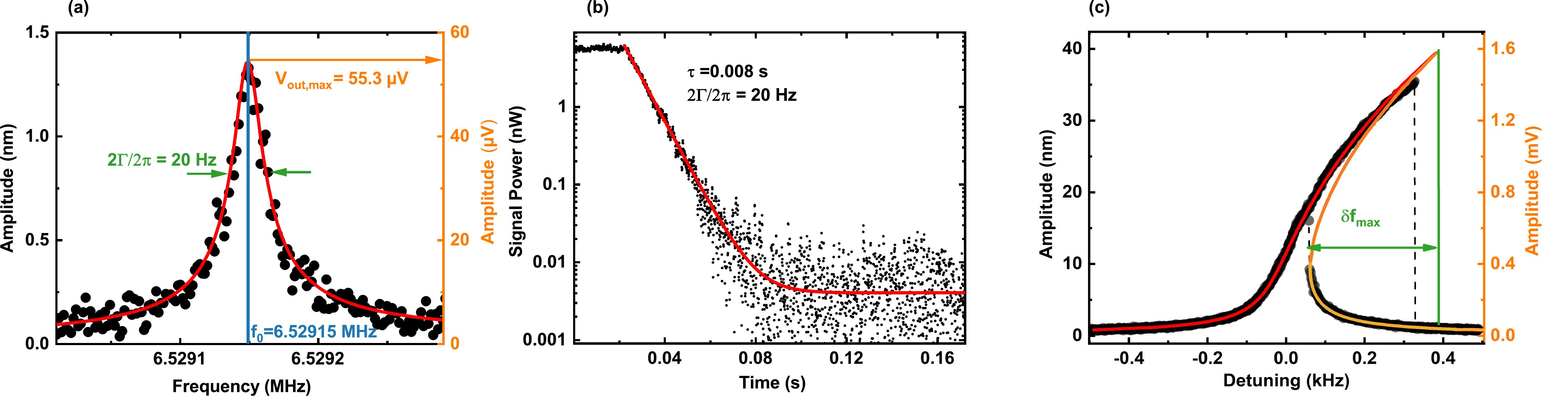}
	\caption{ 
	{\bf a} Linear response measurement (black dots) as well as Lorentzian fit (red line) for 
	 the eigenfrequency and linewidth evaluation as well as for the driving strength calibration. 
	{\bf b}, Ringdown measurement (black dots) along with exponential fit (red line), confirming the linewidth measured in linear response.
	{\bf c} Nonlinear response measurement (black dots) with fit of the Duffing model (red, yellow, and  orange line represents stable high amplitude, 
	 stable low amplitude and unstable solution, respectively). Amplitude and frequency detuning of the backbone point highlighted in green are 
	 used for the amplitude calibration. }
\label{fig:characterization}	 
\end{figure}

In the measurement we cannot access the amplitude $A$ and the force $F$ directly. 
However we know the amplitude of the rf input voltage $V_{\text{in}}$ that drives the capacitor containing the dielectric nanostring and 
we measure the output voltage signal $V_{\text{out}}$ close to the eigenfrequency of the resonator $\omega_0$.
We assume these quantities to be proportional to the driving strength $F$ and to the amplitude $A$, respectively, 
%
%
%
%
%
\begin{equation}
\label{eq:scaling}
A = a \, V_{\text{out}} \,\, \, , \qquad F = b \, V_{\text{in}} \, .
\end{equation}
%
%
%
This allows us to calibrate the driving strength with the benefit of excluding a parameter for further analysis. 
For weak driving power, the resonator is in the linear regime and the vibration amplitude $A$  is given by the well known Lorentzian response
$A^2 = \left(F/2\omega_{0}\right)^2 \, / \, \left( \delta \omega^2+\Gamma^2 \right)$. 
Rewriting the Lorentzian response function with the use of Eq.~(\ref{eq:scaling}) leads to 
%
%
%
%
%
\begin{equation}
\label{eq: Scaling}
V_{\text{out}}^2= \frac{\Gamma^2}{\delta \omega^2+\Gamma^2}\cdot c \cdot V_{\text{in}}^2,
\end{equation}
%
%
%
where the dimensionless parameter $c= {\left( b / a\right)}^2  / (4 \omega_0^2 \Gamma^2)$ is the calibration factor.
At fixed rf input voltage $V_{\text{in}}$ (or drive power),  we fit $V_{\text{out}}$ with the Lorentzian function Eq.~(\ref{eq: Scaling}) 
(see Fig.~\ref{fig:characterization}a), yielding an eigenfrequency $f_{0}=\omega_{0}/2\pi= 6.529$\,MHz 
and a linewidth $2\Gamma/2\pi=20\,$Hz. 
The same linewidth is found by a ringdown measurement (see Fig.~\ref{fig:characterization}b).
Driving the system on resonance ($\delta \omega=0$) gives the maximum output signal $V_{\text{out,max}}$, 
and it follows that $V_{\text{out,max}}^2=  \, c \,   V_{\text{in}}^2$.
This procedure is repeated for different drive voltages $V_{\text{in}}$ (expressed as drive power in the main text). 
The obtained ratios $V_{\text{out,max}}^2/V_{\text{in}}^2$ are fit with the constant $c$. 
We obtained a calibration factor $c = 0.0062(3)$.

%
%
%
%
%
%
%

%
%
%
\subsection{Duffing nonlinearity parameter}
From Eq.~(\ref{eq:Amp}) one can express the detuning  $\delta \omega$ in terms of the amplitude $A$ of the forced vibrations on the high-amplitude branch.
Using the scaling relations Eq.~(\ref{eq:scaling}), we can write 
%
%
%
%
%
%
\begin{equation}
\label{eq:fitting_duffing}
\frac{\delta \omega}{\Gamma} 
= 
\frac{3\tilde{\gamma}}{8\omega_{0}\Gamma} \, V_{\text{out}}^2
\, - \,
\sqrt{\ c \, \left( \frac{V_{\text{in}}^2}{V_{\text{out}}^2} \right)  - 1 },
\end{equation}
%
%
%
%
where we have introduced the Duffing parameter $\tilde{\gamma} = \gamma \cdot a^2$ expressed in units of $[1/ (\mbox{V}^2 \mbox{s}^2)]$. 
In the nonlinear regime, we fit the nonlinear amplitude response function  by plotting $\delta \omega$ 
as a function of the r.h.s. of Eq.~(\ref{eq:fitting_duffing}) with the high amplitude branch for $V_{\text{out}}$ and using $\tilde{\gamma}$ as the single fitting parameter.
For the power $P= -31\, \mbox{dBm}$, which is used to obtain the results shown in Fig.~1b of the main text, and for $P= -30 \, \mbox{dBm}$    shown in Fig.~\ref{fig:characterization}c 	  
we find the scaled Duffing parameter to be $\tilde{\gamma} = 9.28 \cdot 10^{16} \, \mbox{V}^{-2} \, \mbox{s}^{-2}$. 
Alternatively and independently, $\tilde\gamma$ can be determined from the relation between the maximal value of the amplitude of the nanoresonator measured in volts, $V_{\rm out,\max}$, and the frequency detuning of the drive $\delta\omega_{\max}$ at which this maximal value is reached on the upper branch of the response curve, see Fig. S3c, i.e., from the so-called Duffing backbone curve

\begin{equation}
\label{eq:backbonecalib2_EW}
\tilde\gamma V^2_{\text{out, max}} = \frac{8}{3} \omega_{0} \delta \omega_{\text{max}} \, ,
\end{equation} 
The value of $\tilde{\gamma}  \simeq 9.3 \cdot 10^{16} \, \mbox{V}^{-2} \, \mbox{s}^{-2} $ 
obtained from  the  measurements based on Eq.~(\ref{eq:backbonecalib2_EW}) for five values  
of the input power coincided with the value given above, providing an independent proof of the 
applicability of the Duffing model in the studied parameter range.

With the measured calibration factor $c$, the scaled Duffing parameter $\tilde\gamma$,  the eigenfrequency, and the decay rate of the mode, we have fully determined the parameters of the theoretical model, Eq.(1) of the main text, except for the scaling $a$.

%
%
%
%
\subsection{Amplitude conversion}
To find the scaling $a$ and thus to convert the measured displacement amplitude in volts $V_{\rm out}$ into the amplitude $A$ in meters  we use the procedure described in Ref.~\onlinecite{Pernpeintner:2014ks}. It relies on the fact that our nanoresonator can be reliably modeled by a string. For a string, the value of $\gamma$ in the units of ${\rm (m\cdot s)}^{-2}$ can be obtained using the nonlinear dependence of its bending-induced elongation on the vibration amplitude. This effect is often referred to as geometric nonlinearity~\cite{Schmid2016}. For the nanoresonators of the type studied in this paper, this is the major nonlinearity mechanism. It yields the theoretical value
\begin{equation}
\label{eq:calculated_gamma}
\gamma_{\text{th}} \, = \,  \pi^4 \frac{E}{4l^4 \rho} \approx 1.54 \cdot 10^{26}\,\text{m}^{-2}\text{s}^{-2} \, ,
\end{equation}
where $E= 160 \,\, \mbox{GPa}$ \cite{Unterreithmeier2010} is the Young's modulus and $\rho= 2800 \,  \mbox{kg} / \mbox{m}^3 $ is the mass density \cite{el-Hak2002}.

Using the relation between $\tilde\gamma$ and $\gamma$, we obtain
	\begin{equation}
	\label{eq:a}
			a  =  (\tilde\gamma/\gamma)^{1/2} \approx (\tilde\gamma/\gamma_{\rm th})^{1/2} \simeq 2.45 \times 10^{-5}{\rm m/V} \, .
		\end{equation}

Finally, we note that the amplitude conversion factor $a$ can alternatively be determined from the thermomechanical fluctuations of the resonator, provided the thermal motion of the undriven nanostring is resolved. This is not the case in the presented experiment, such that the amplitude conversion of our data relies on the nonlinear method described above.

%
%
%
%
%
%
\subsection{Additional data: Power spectra as a function of the drive power for fixed detuning near the critical switching point}
%
%
%
%
%
%
%
%
\begin{figure}[b]
	\includegraphics[width=\columnwidth]{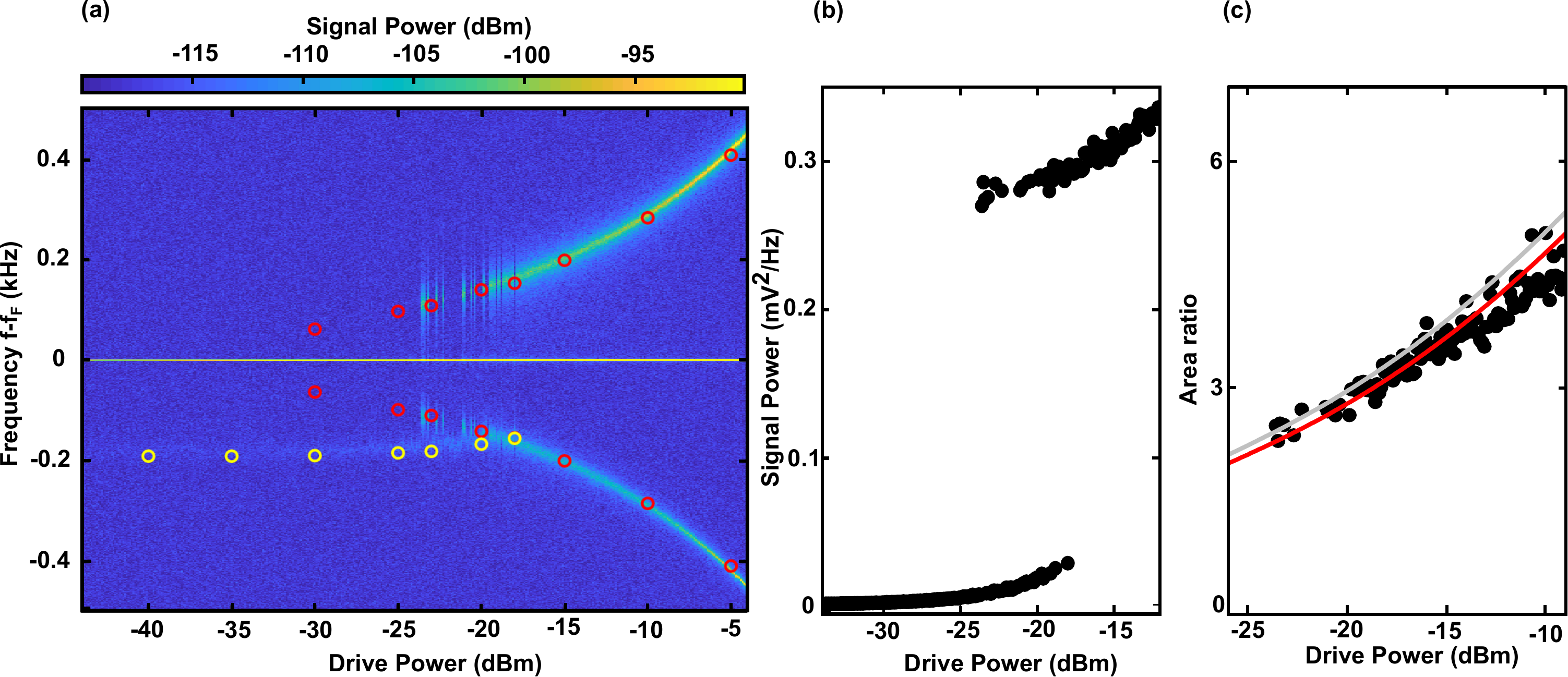}
	\caption{ Power spectra as a function of the drive power for a fixed detuning of 190\,Hz.
		{\bf (a)}  Color coded power spectra showing the satellite splitting and intensity as a function of the drive power. 
		Red (yellow) open circles denote theoretically calculated satellite positions for the high (low) amplitude solution 
		in the weak-damping approximation, that applies only where the   satellite peaks   are well-resolved.
		{\bf (b)}  Extracted amplitude of the response at the drive tone as a function of the drive power.
		{\bf (c)} Ratio of the satellite areas when the resonator is in the high amplitude state. 
		Gray (red) line shows the theoretical prediction in the weak damping approximation (more general linearized approximation taking into account the overlap of the satellite peaks).	
	}
	\label{figure:spectra_detuning}
\end{figure}
%
%
%

As an additional measurement we apply a fixed detuning of 190\,Hz to the actuation frequency 
(close to the measured critical switching point) and sweep the drive power from  -45\,dBm up to -4\,dBm, 
comparable to the measurement under resonant drive in the main text (Fig.\,2a). 
The resulting power spectra as a function of drive power are shown in Fig.~\ref{figure:spectra_detuning}a. 
In contrast to the data obtained for a resonant drive, we observe an additional, single lower frequency satellite. 
This satellite is observed for a weak drive (below -30\,dBm). 
For a strong drive (above -15\,dBm), 
two symmetrical satellites are observed, similarly to the data obtained for a resonant drive. 
In the regime in between, each power spectrum shows either the former or the latter satellites, such that both satellite branches are visible. 
By calculating the expected satellite splitting, we can assign the single satellite to the low amplitude state of the Duffing resonator 
(yellow circles) and the two symmetric satellites to the high amplitude state (red circles).
In the intermediate drive range, the Duffing resonator is in the bistable regime and depending on the initial conditions 
the system chooses either the high or the low amplitude solution, 
consequently we see a switching behaviour between the two possible satellite branches.
This is particularly apparent from Fig.~\ref{figure:spectra_detuning}b, 
which plots the amplitude response at the drive tone as a function of the drive power, as in Fig.~4b of the main text. 
For a weak drive (below -30\,dBm), the resonator is always in the low amplitude state, whereas for a large drive (above -15\,dBm), 
the system is only found in the high amplitude state. 
In the intermediate region, the amplitudes jumps back and forth, leading to the behavior observed for the satellite branches.
As before, the area of the satellites can be extracted by a Lorentzian fit and the calculated ratio of the areas is plotted in Fig.~\ref{figure:spectra_detuning}c. 
We can only report a ratio for the high amplitude solution, as only one of the two satellites of the low amplitude solution is resolved in the experiment.
Again the calculated area ratio is in a good agreement with the theory, where the red line corresponds to the theoretical calculation including the overlap of the two peaks and the gray line to the additional weak damping approximation.

%
%
%
%
%
%

\subsection{Additional data: Frequency fluctuations and direct homodyne detection of the noise quadratures}
%
%
%
%
%
%
%
%
%
%
%
\begin{figure}[b]
\includegraphics[width=0.8\columnwidth]{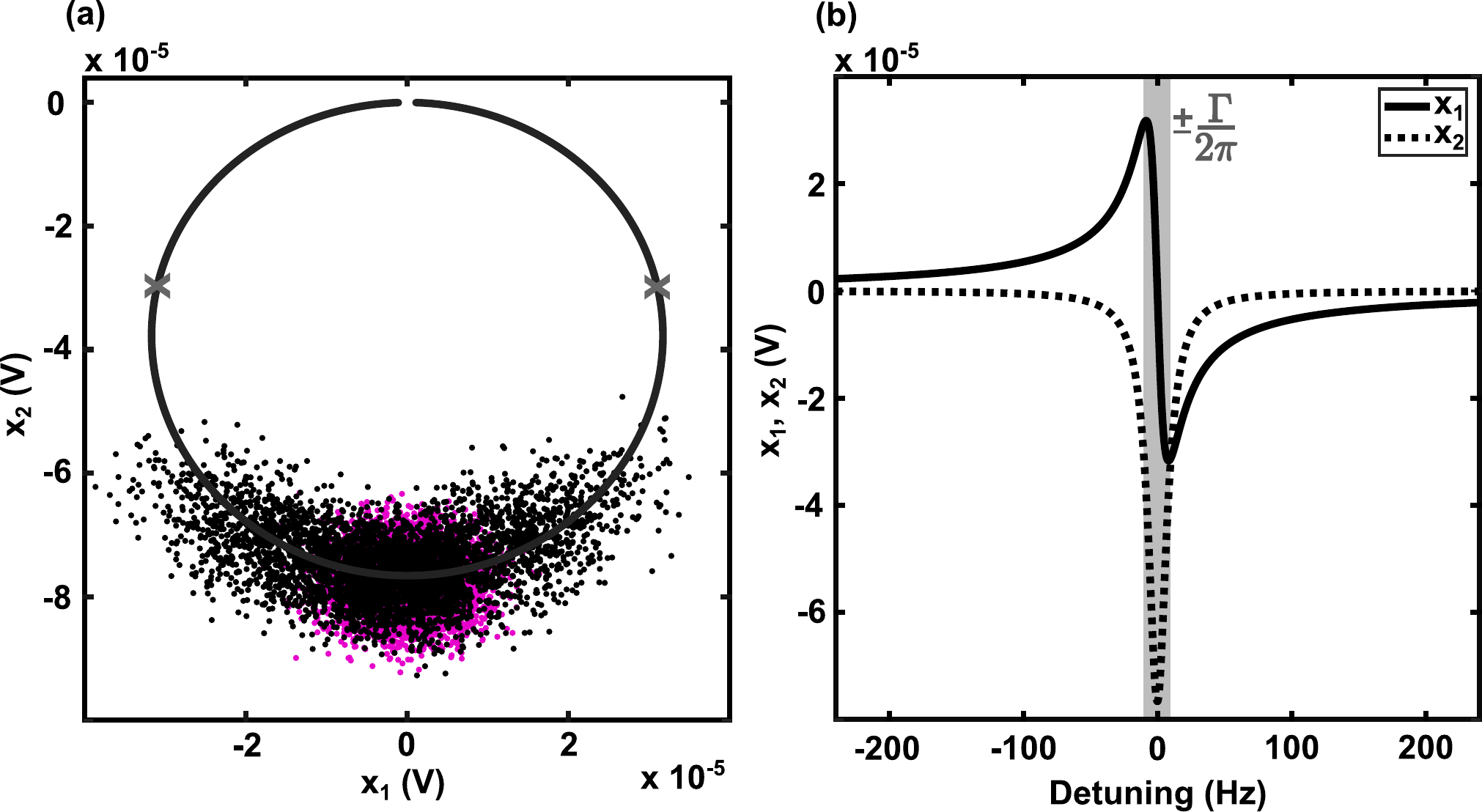}
\caption{In-phase and quadrature component of the response of the driven high Q resonator. {\bf (a)} Homodyne measurement of the in-phase and quadrature component of the driven resonator's response. The black dots show the noise distribution for a resonant drive in the linear regime. The pink dots show the noise distribution when repeating the measurement under active frequency stabilization using a phase-locked loop. The solid line illustrates the phase space representation of the full resonance curve, and the gray crosses mark a detuning of $\pm\Gamma/(2\pi)$ from resonance. {\bf (b)} Calculated in-phase and quadrature components $x_1$ and $x_2$ of the response of the driven resonator using the same parameters. The gray shaded region also marks a detuning of $\pm\Gamma/(2\pi)$.}
\label{fig:banana}
\end{figure}
%
%
%

In an attempt to complement and validate the spectral measurements of the thermal squeezing, we perform a homodyne measurement to resolve the in-phase and quadrature component of the thermal fluctuations around the stable state of forced vibrations. To this end, the fast lock-in amplifier is employed as described in section II.A. In order to capture the thermomechanical noise
in the linear regime, the lock-in bandwidth needs to exceed the linewidth of the resonator, $2\Gamma/(2\pi)$. We employ a lock-in bandwidth of $50$\,Hz with a fourth order filter. For a constant drive frequency and power, the in-phase and quadrature signal is captured over a time interval of $20$\,s to sample the underlying noise distribution in approx. $5,000$ points.

Figure~\ref{fig:banana}a shows the measured phase-dependent noise of the resonantly driven resonator ($f_F = f_0$) for a drive power of $-56$\,dBm in the linear response regime. Theoretically, we expect to find a circularly symmetric noise distribution centered around a point which is displaced from the origin by the drive. For the case of thermal noise, we expect a Gaussian distribution with a variance of $77 \cdot 10^{-12}$\,V$^2$. It is immediately apparent that the data (black dots) is not in agreement with this expectation. Rather than a circular distribution, we find a 'banana' following the black line which corresponds to the phase space representation of a frequency sweep through the resonance. This indicates that the system does not stay on resonance for the duration of the measurement. The distored shape of the noise distribution is thus attributed to frequency fluctuations of the resonator, as discussed in Ref.~\cite{Fong2012}. 
For the case of a strongly resonantly driven high Q resonator, these frequency fluctuations translate into fluctuations of the quadrature $x_1$ which are much larger than the thermomechanical fluctuations and thus obviate their experimental determination in a 
homodyne measurement.

A more thorough understanding of the impact of frequency fluctuations can be obtained from Fig.~\ref{fig:banana}b which illustrates the in-phase and quadrature components $x_1$ and $x_2$ of the response of the driven resonator in the linear regime. It is immediately apparent that near zero detuning ($f_F = f_0$) the in-phase component $x_1$ scales linearly with the detuning. More precisely, $\delta x_1 \approx   \delta \Omega F / (2  \Gamma^2 \omega_0)$ around $\omega_F=\omega_0$ for frequency fluctuations of amplitude $\delta\Omega$. This implies that for weakly damped resonators, even weak frequency fluctuations lead to a large broadening of the measured $x_1$. 
The quadrature component $x_2$ is only quadratically sensitive to frequency fluctuations.
On the other hand, as discussed in the main text, the frequency fluctuations only lead to a negligible broadening of the power spectrum as long as they are weak ($<\Gamma$). Therefore the spectral measurement is better suited to characterize the squeezing of high Q resonators.

In principle, the eigenfrequency fluctuations of the resonator can be tracked and compensated with the lock-in amplifier's integrated phase-locked loop (PLL) option. This indeed removes the distorted shape of the noise distribution (see pink dots in Fig.~\ref{fig:banana}a). However, the feedback controller produces additional noise which is injected into the resonator~\cite{Roy2018}. Therefore, the resulting circularly symmetric noise distribution is still not suitable to extract the thermomechanical fluctuations of the resonator. 

That being said, squeezed quadratures of high Q resonators elude a homodyne measurement. 
As a matter of fact, to the best of our knowledge, homodyne detection of squeezing of a high Q resonator mode has not been accomplished, neither in the mechanical nor in the microwave domain.
Therefore, the presented method of inferring squeezing from the sideband-resolved power spectrum gives access to the noise quadratures in a regime where they cannot be directly determined.

%
	

%